\documentclass{aa}

\usepackage{graphicx}
\usepackage{txfonts}
\usepackage{subcaption} %
\usepackage{lscape} %
\usepackage{placeins} %

\usepackage{booktabs}
\usepackage{multirow,xcolor}

\usepackage{xcolor}
\AtBeginDocument{\nolinenumbers\let\linenumbers\nolinenumbers}

\begin{document}
   \title{Anisotropic redshift distributions in photometric galaxy clustering and their cosmological impact}
   \titlerunning{Anisotropic redshift distributions in galaxy clustering}
   \authorrunning{Z. Zhang et al.}

   \author{Zekang Zhang \inst{1}
   \thanks{\email{zekang.zhang@physik.lmu.de}}
   \and Yunhe Wang \inst{1} \and Daniel Gruen \inst{1,2} \and Hui Kong \inst{1,3} \and \\
    Luca Tortorelli \inst{1} \and Silvan Fischbacher\inst{4} \and Ziang Yan\inst{5,6}}

   \institute{Universitäts-Sternwarte, Fakultät für Physik, Ludwig-Maximilians Universität München, Scheinerstr. 1, 81679 München, Germany \\ %
   \and Excellence Cluster ORIGINS, Boltzmannstr. 2, 85748 Garching, Germany \\ %
   \and
   {Institut de Física d’Altes Energies (IFAE), The Barcelona Institute of Science and Technology,
Campus UAB, 08193 Bellaterra, Barcelona, Spain} \\
   \and
   Institute for Particle Physics and Astrophysics, ETH Zurich, Wolfgang-Pauli-Strasse 27, CH-8093 Zurich, Switzerland\\
   \and
   Graduate School of Science, Nagoya University, Furocho, Chikusa-ku, Nagoya, Aichi, 464-8602, Japan\\
   \and
   Kobayashi-Maskawa Institute for the Origin of Particles and the Universe (KMI), Nagoya University, Nagoya, 464-8602, Japan
   }

   \date{Received September 30, 20XX}

   \abstract
   {
      {Photometric galaxy clustering is a major probe of large-scale structure.}
      Its interpretation rests on accurate treatment of the redshift distribution, $n(z)$, of
      the galaxy sample. Standard analyses assume the $n(z)$ of a selected lens galaxy sample
      to be isotropic, whereas observational
      systematics imprint a variation of it across the sky.
   }
   {
      The cosmological impact of this $n(z)$ anisotropy has not yet been studied
      end-to-end at survey-level realism. We quantify how it biases the angular
      two-point correlations $w(\theta)$, the internal consistency of a cosmological analysis of these data, and the inferred cosmological parameters.
   }
   {
      We develop an analytical formalism of angular galaxy clustering under spatially varying $n(z)$,
      and forward-model the full observation chain: from realistic maps of observational systematics,
      a machine-learning emulated galaxy selection function, to tomographic binning.
      We compute the resulting observed $w(\theta)$ and propagate the impact of anisotropic $n(z)$
      {to a $2\times2$pt analysis for a KiDS-like survey.}
      We extend the analysis to configurations that adopt the
      larger DES- and LSST-sized tiles together with each survey's area and number density.
   }
   {
      For our fiducial magnitude-limited sample, modeling $w(\theta)$ with a
      single global $\bar{n}(z)$ underestimates the true clustering signal by up to $\sim$10\% at
      small scales in the highest tomographic bin, falling to a few percent at
      intermediate redshift and vanishing at low redshift. Propagated to cosmology,
      the parameter shifts stay below $1\sigma$ for the KiDS-like and DES-like setups,
      but for the LSST-like setup $\sigma_8$ is biased low by $1.7\sigma$ and
      $\Omega_m$ high by $1.4\sigma$ along their degeneracy. The contamination is preferentially absorbed by the
      high-redshift galaxy bias, which shifts by up to $\sim$4$\sigma$.
      {The anisotropy also affects whether the probes are consistent enough
      to combine:} a calibrated posterior-predictive internal-consistency test finds
      the contaminated angular clustering improbable given the joint cosmic-shear and galaxy--galaxy-lensing
      constraints (median calibrated $\tilde p\approx8\times10^{-4}$ under LSST-like noise).
   }
   {
      {Anisotropic redshift distributions have a small but non-negligible effect
      in our KiDS-like and DES-like configurations and a significant effect in our
      LSST-like configuration, which retains KiDS-like selection and observing-condition variation.} Because our
      forward model is predictive and independent of cosmology, the same machinery
      {that quantifies the effect could be inverted to mitigate it given a realistic selection function and galaxy population model.}
   }


   \maketitle

   \section{Introduction}

   {The clustering of galaxies is one of the major probes of large-scale
   structure and has been used extensively to test the cosmological model.
   Photometric surveys measure galaxy clustering using broadband imaging.
   Due to limited spectral information, they recover accurate
   angular positions but only uncertain redshifts and thus line-of-sight distances. The clustering of these positions is therefore usually quantified by the angular two-point correlation function}
   $w(\theta)$, the excess probability over random of finding two galaxies
   separated by an angle $\theta$ on the sky. Early angular
   clustering measurements from the Sloan Digital Sky Survey \citep{Connolly2002}
   established the approach, and {modern photometric surveys such as the Dark
   Energy Survey (DES, \citealt{ElvinPoole2018})} yield competitive constraints on the
   matter density $\Omega_{\rm m}$ and the amplitude of density fluctuations
   $\sigma_8$ from probes that include angular galaxy clustering, complementing those from the cosmic microwave background and
   baryon acoustic oscillations.

   Galaxies are biased tracers: their clustering can be related to that of the
   underlying dark matter, differing by an amplitude that acts multiplicatively in
   the linear regime --- an effect commonly referred to as galaxy bias. On linear scales the
   clustering {signal} scales as $w(\theta)\propto b^2(z)\,\sigma_8^2$, so the
   bias is strongly degenerate with fundamental cosmological parameters.
   State-of-the-art analyses therefore combine
   multiple probes, in particular galaxy clustering with gravitational
   lensing, the latter being especially valuable as it probes dark matter
   directly. Such joint constraints have been applied successfully and have
   become standard in modern observations {(e.g., the DES $3\times2$pt
   analysis \citealt{DESY62026}).}
   {The upcoming Vera C. Rubin Observatory Legacy Survey of Space and Time
   \citep[LSST;][]{IvezicLSST2019} and Euclid
   \citep{LaureijsEuclidReport2011} will map billions of galaxies over
   unprecedented areas, driving the systematic uncertainty budget down to
   sub-percent levels.}

   Because photometric surveys do not resolve individual distances, the central
   challenge is to characterize the redshift distribution $n(z)$ of the tracer
   ensemble, which sets the projection of the three-dimensional galaxy field
   onto the sky. This distribution must be estimated to very high accuracy in order to
   avoid biasing the cosmological inference: requirements on the
   per-bin mean redshift are at the level of $|\Delta z| \lesssim {0.003}(1+z)$ for
   {the LSST DESC Year-10 lens sample} \citep{LSSTSRD2018}.

   {Besides the characterization of the redshift distribution, another central problem in galaxy clustering is spatial systematics, whereby observational variations across the sky translate into directional variations in galaxy selection. For example, the Galactic foreground extinguishes and reddens the light from galaxies and is itself spatially distributed, while observing conditions, whether atmospheric or instrumental, vary with time and position and coherently affect data quality. These systematics introduce spatial variations in the sample density that can enter the measurement as a false signal of cosmic structure.
   }

   {A variety of approaches has been proposed and applied to survey data
   to correct spatial variations in galaxy number density caused by observational selection.}
   These include Template Subtraction \citep{Ross2011, Ho2012, ElvinPoole2018},
   Mode Projection \citep{Leistedt2014} and Iterative Systematics Decontamination
   \citep{RodriguezMonroy2022,DESY6Clustering2026}.
   Machine-learning-based methods have also been
   proposed to map non-linear relationships between survey property maps and galaxy
   density \citep{Rezaie2020,YanClustering2025}. Broadly speaking, these can all
   be viewed as variants of a regression method, in which property maps are derived
   externally and used to correct the data map \citep{WeaverdyckHuterer2021}. The corrections of the above methods
   have been validated extensively on mock
   catalogs and applied to data, demonstrating that uncorrected density
   fluctuations can bias cosmological parameters by several $\sigma$ if left
   untreated \citep{RodriguezMonroy2022,YanClustering2025}.

   A direction left largely unexplored by almost all of the methods above is
   the anisotropic distribution of galaxy properties, not just their number density.
   Although clustering formally probes only the angular positions of
   galaxies, the observed signal depends not only on the fraction of galaxies
   selected at each point on the sky, but also on \emph{which} galaxies those
   are. Spatially varying selection modulates the population
   composition of the sample, not just its number density. This composition
   enters the model through two redshift-dependent ingredients: the redshift
   distribution $n(z)$ and the galaxy bias $b(z)$.
   The redshift distribution of a sample
   depends on the selection function, which in turn varies with the
   observational systematics. The same holds for galaxy bias. Standard
   modeling assumes both quantities to be isotropic, {i.e.\, their expected values have no angular dependence from observational selection.}
   {Stage-III and earlier survey analyses adopted this assumption without
   closely testing it, on the grounds that departures from isotropy would have
   a negligible impact within their uncertainty budgets.} {The tighter uncertainty budgets of Stage-IV surveys motivate testing
   the impact of departures from isotropy.}
   \citet{LizancosSpatial2023} developed {a theoretical formalism} for
   this problem, showing that anisotropy in the redshift distribution induces
   a mode-coupling effect that adds spurious power to auto-correlations such
   as galaxy clustering and cosmic shear, while certain cross-correlations,
   such as {cross-correlations with CMB lensing}, remain almost entirely immune.
   \citet{Kong2026} termed this class of effects ``sub-sample systematics'',
   showing that even when the overall sample density is correctly calibrated,
   spatial variation in $n(z)$ and $b(z)$ enhances the
   auto-correlation amplitude in a way that standard imaging-systematics
   mitigation cannot correct. They derived analytic expressions for this
   effect and discussed forward-modeling approaches for its
   estimation. Most closely related empirically,
   \citet{HangLSST2024} propagated
   spatially varying LSST depth through a realistic photometry and
   photometric-redshift pipeline to quantify how the tomographic
   redshift distributions vary across the footprint, forecasting
   via a Fisher analysis that galaxy clustering is the probe most
   susceptible to this effect.

   {We extend these studies by propagating the effect to cosmological parameter
   inference. We forward-model how realistic observing-condition maps affect selection,
   tomographic binning, and $w(\theta)$, and quantify the resulting parameter bias.}
   We first derive an analytical
    model of anisotropic galaxy clustering, expressing the contamination {of}
    $w(\theta)$ in terms of the spatial perturbation of the redshift
    distribution. We then realize this model through survey-realistic forward
    modeling. {To this end we map {observing conditions},} via an ML-based selection
    emulator, to the position-dependent $n(z)$ and the resulting additive bias
    on $w(\theta)$. {The forward model predicts selection-induced variations in $n(z,\boldsymbol{\theta})$
    independently of the clustering cosmology. Combined with a model of the matter
    correlation function, these predictions could be used to correct $w(\theta)$,
    given a sufficiently realistic model of the survey selection function.} We validate
    the model on simulated footprints and propagate the residual bias to
    cosmological parameter shifts in a $2\times2$pt likelihood across
    KiDS-, DES-, and LSST-like configurations.

   The paper is organized as follows. In Sect.~\ref{sec:formalism} we derive the
   formalism for anisotropic angular clustering. Sect.~\ref{mocks}
   describes the construction of spatially varying systematics maps and the
   selection emulator. We present our main results on the clustering enhancement
   in Sect.~\ref{sec:main_results}, and propagate the effect to cosmological
   parameter inference in Sect.~\ref{sec:cosmo}. We conclude in
   Sect.~\ref{sec:conclusions}.

   \section{Formalism}
   \label{sec:formalism}

   {The angular two-point correlation function (2PCF) $w(\theta)$ is a core statistical tool in
   cosmology, quantifying the excess probability, relative to a random
   distribution, of finding a pair of galaxies at a given
   angular separation.} As the configuration-space counterpart of the power spectrum,
   it captures the Gaussian information of the matter field.

   Most cosmic mass is in the form of dark matter, which is not directly observable.
   Luminous tracers such as galaxies are therefore used as indirect, biased
   probes of the underlying density field. In spectroscopic surveys their full
   three-dimensional positions can be reconstructed, but in photometric surveys
   only broadband imaging is available and individual line-of-sight distances
   are uncertain. It is then most practical to measure angular correlations of
   galaxy fields projected along the line of sight, with the projection set by the
   galaxy redshift distribution $n(z,\boldsymbol{\theta})$, where
   $\boldsymbol{\theta}$ denotes sky position. Consistency between data and
   model requires that the same $n(z,\boldsymbol{\theta})$ enter both.

   Standard analyses assume that the selection is statistically uniform across
   the footprint, so that $n(z,\boldsymbol{\theta}) \equiv n(z)$ is spatially
   isotropic, depending only on $z$. In practice, however,
   data are affected by spatially varying observational systematics that break
   {this assumption}. In this section, we derive the corresponding formalism for the case when
   the isotropy assumption is violated. Similar {perturbative} forms of spatial
   $n(z,\theta)$ are seen in works including {\cite{LizancosSpatial2023} and \cite{Kong2026}}.

   We begin by writing the projected galaxy overdensity as a line-of-sight
   integral over the 3D galaxy field,
   \begin{equation}
      1+\delta^{\mathrm{2D}}(\boldsymbol{\theta}) = \int \mathrm{d}z\, n(z,\boldsymbol
      {\theta}) \left[1+\delta_{g}^{\mathrm{3D}}(z,\boldsymbol{\theta})\right].
   \end{equation}
   Here $n(z,\boldsymbol{\theta})$ is the \emph{expected} redshift distribution
   of galaxies selected at sky position $\boldsymbol{\theta}$, set by the local
   observing conditions and intrinsic galaxy properties, with no modulation by the matter density field. It acts as the line-of-sight kernel that sets the relative weight of
   each redshift in the projected sample. $\delta_{g}^{\mathrm{3D}}$ denotes the
   3D galaxy overdensity field, namely the overdensity field of discrete tracers
   (galaxies) that are biased with respect to the underlying dark matter field
   $\delta_{m}^{\mathrm{3D}}$. The subscript $g$ in $\delta^{\mathrm{2D}}$ is
   omitted for brevity. {The two factors separate cleanly: {the clustering of galaxies} in
   matter overdensities makes the observed counts anisotropic on the sky, but
   that anisotropy is the clustering signal itself and is carried entirely by
   $\delta_{g}^{\mathrm{3D}}$, while the kernel $n(z,\boldsymbol{\theta})$ varies
   across the sky only through selection.} Folding any of the clustering
   anisotropy into $n(z,\boldsymbol{\theta})$ would produce a non-zero
   $w(\theta)$ even in the absence of matter clustering.

   On the angular scales considered here, we assume linear, deterministic
   galaxy bias,
   \begin{equation}
      \delta_{g}^{\mathrm{3D}}(z,\boldsymbol{\theta})
      = b(z,\boldsymbol{\theta})\,\delta_{m}^{\mathrm{3D}}(z,\boldsymbol{\theta}),
   \end{equation}
   where $b(z,\boldsymbol{\theta})$ is the linear-bias coefficient. Standard
   models take $b$ to depend only on redshift. {However,} the same spatially
   varying selection criteria that modulate $n(z,\boldsymbol{\theta})$ also
   modulate the sample composition: galaxies of different color, type, or
   luminosity carry distinct intrinsic biases, so observing-condition gradients
   imprint a residual angular dependence on the effective bias. {A similar spatial perturbation can therefore be introduced for the galaxy bias. Nevertheless, in what follows we hold the galaxy bias isotropic, $b(z,\boldsymbol{\theta})=\bar b(z)$, and defer its spatial variation to future work.}

   We parameterize the redshift distribution as a perturbation around its
   global mean,
   \begin{equation}
      \label{perturbations} n(z,\boldsymbol{\theta}) = \bar n(z)\left[1+\epsilon(z,\boldsymbol{\theta})\right].
   \end{equation}

   The normalization of the global redshift distribution requires
   \begin{equation}
      \int \mathrm{d}z\,\bar n(z)=1.
   \end{equation}

   Because the 2PCF probes the number of galaxy pairs, density fluctuations
   induced by spatial selection effects are a dominant source of contamination.
   If this density modulation is not corrected for, one has
   \begin{equation}
      \int \mathrm{d}z\,\bar n(z)\,\epsilon(z,\boldsymbol{\theta}) \equiv 1-s(\boldsymbol
      {\theta}) \neq 0.
   \end{equation}
   Here $s(\boldsymbol{\theta})$ denotes the spatial selection function,
   capturing the angular modulation of galaxy number density induced by
   observational systematics. This contamination is addressed in standard treatments by
   applying selection weights to the galaxy sample \citep[e.g.,][]{ElvinPoole2018}, or by using
   structured random catalogs in the Landy--Szalay estimator \citep[e.g.,][]{Suchyta2016, Everett2022}. In the
   main body of this work, we assume this mitigation is already successful and
   set $s(\boldsymbol{\theta})=1$.

   {The global redshift distribution can be defined as an area-weighted
   or a count-weighted average. For an area-weighted $\bar{n}(z)$, every equal-area
   sky element contributes equally, so the angular average of $\epsilon$
   vanishes by construction,}
   \begin{equation}
      \langle \epsilon(z,\boldsymbol{\theta}) \rangle = 0. \label{weighted_global_nz}
   \end{equation}
   Throughout this paper,
   $\langle\cdot\rangle$ denotes an angular average over sky positions $\boldsymbol{\theta}$.
   {By default, we construct the global redshift distribution from the selected
   galaxy sample without inverse-density weights, giving a count-weighted
   $\bar{n}(z)$. The difference between count-weighted and area-weighted definitions
   is demonstrated in Appendix~\ref{sec:impact_n_def}.}

   We now expand the angular two-point correlation of the projected
   tracer overdensity:
   \begin{align}
      \bigl\langle \delta^{\mathrm{2D}}(\boldsymbol{\theta})\,\delta^{\mathrm{2D}}(\boldsymbol{\theta}') \bigr\rangle
        & = \int \mathrm{d}z_{1}\int \mathrm{d}z_{2}\,
        \bigl\langle n_{1}\,\delta_{1}\,n_{2}'\,\delta_{2}' \bigr\rangle,
   \end{align}
   where we have introduced the shorthand
   $n_{i}\equiv n(z_{i},\boldsymbol{\theta})$,
   $n_{i}'\equiv n(z_{i},\boldsymbol{\theta}')$, and similarly
   $\delta_{i}\equiv\delta_{g}^{\mathrm{3D}}(z_{i},\boldsymbol{\theta})$,
   $\delta_{i}'\equiv\delta_{g}^{\mathrm{3D}}(z_{i},\boldsymbol{\theta}')$.
   We reduce this expression in two steps.

   Firstly, $n(z,\boldsymbol{\theta})$ is set by observational systematics and
   intrinsic galaxy properties, and is therefore statistically independent of
   the underlying matter field at leading order. The four-point average then
   factorizes,
   \begin{equation}
      \bigl\langle n_{1}\,\delta_{1}\,n_{2}'\,\delta_{2}' \bigr\rangle
      \;\approx\;
      \bigl\langle n_{1}\,n_{2}' \bigr\rangle\,
      \bigl\langle \delta_{1}\,\delta_{2}' \bigr\rangle.
   \end{equation}
   The leading correction {to this} comes from blending: detection efficiency
   depends on neighboring objects, inducing a residual coupling between
   $n(z,\boldsymbol{\theta})$ and the local matter density.
   We neglect this effect in this work.

   Secondly, for a broad redshift kernel, the matter correlation length is much smaller than the kernel width. Consequently, the contribution to the projected correlation function is strongly localized around $z_1\simeq z_2$ in the radial integration \citep{Limber1953}:
   \begin{equation}
      \bigl\langle \delta_{1}\,\delta_{2}' \bigr\rangle
      \;\approx\;
      \delta_{D}(z_{1}-z_{2})\,\langle\delta\,\delta'\rangle,
   \end{equation}
   where
   $\langle\delta\,\delta'\rangle\equiv
    \bigl\langle \delta_{g}^{\mathrm{3D}}(z,\boldsymbol{\theta})\,
    \delta_{g}^{\mathrm{3D}}(z,\boldsymbol{\theta}') \bigr\rangle$
   denotes the equal-redshift {galaxy correlator}, into which we have absorbed
   the line-of-sight volume element $\mathrm{d}\chi/\mathrm{d}z=c/H(z)$ for
   notational brevity.

   Combining the two steps and performing the $z_{2}$ integral collapses
   the double integral to
   \begin{equation}
      \bigl\langle \delta^{\mathrm{2D}}(\boldsymbol{\theta})\,\delta^{\mathrm{2D}}(\boldsymbol{\theta}') \bigr\rangle
      = \int \mathrm{d}z\, \bigl\langle n(z,\boldsymbol{\theta})\,n(z,\boldsymbol{\theta}') \bigr\rangle\,\langle\delta\,\delta'\rangle.
   \end{equation}

   Applying Eq.~\ref{perturbations}, we obtain
   \begin{align}
       & \bigl\langle \delta^{\mathrm{2D}}(\boldsymbol{\theta})\,\delta^{\mathrm{2D}}(\boldsymbol{\theta}') \bigr\rangle \notag                                                    \\
       & \quad= \int \mathrm{d}z\,\bar n^{2}(z)\,\langle\delta\,\delta'\rangle \label{term1}                                                                                       \\
       & \quad\quad+ 2\int \mathrm{d}z\,\bar n^{2}(z)\,\langle \epsilon(z,\boldsymbol{\theta})\rangle\,\langle\delta\,\delta'\rangle \label{term2}                                 \\
       & \quad\quad+ \int \mathrm{d}z\,\bar n^{2}(z)\,\langle \epsilon(z,\boldsymbol{\theta})\epsilon(z,\boldsymbol{\theta}')\rangle\,\langle\delta\,\delta'\rangle. \label{term3}
   \end{align}
   Three terms appear. {Eq. \ref{term1}} is the standard contribution under an isotropic
   redshift distribution. {Eq. \ref{term2}}, which we refer to as the \emph{shift term}, appears when Eq.~\ref{weighted_global_nz} does
   not hold. {For a count-weighted global redshift distribution,
   $\langle\epsilon(z,\boldsymbol{\theta})\rangle$ can be nonzero because pixels
   with a larger number of selected galaxies contribute more to $\bar{n}(z)$.}
   It increases with stronger angular dependence in sample selection. {Eq. \ref{term3}}, which we call the \emph{clustering term}, arises from spatial variation of the normalized redshift distribution
   and is the main focus of this work.

   \section{Methodology}
   \label{mocks}

     \begin{figure}
      \centering
      \includegraphics[width=\columnwidth]{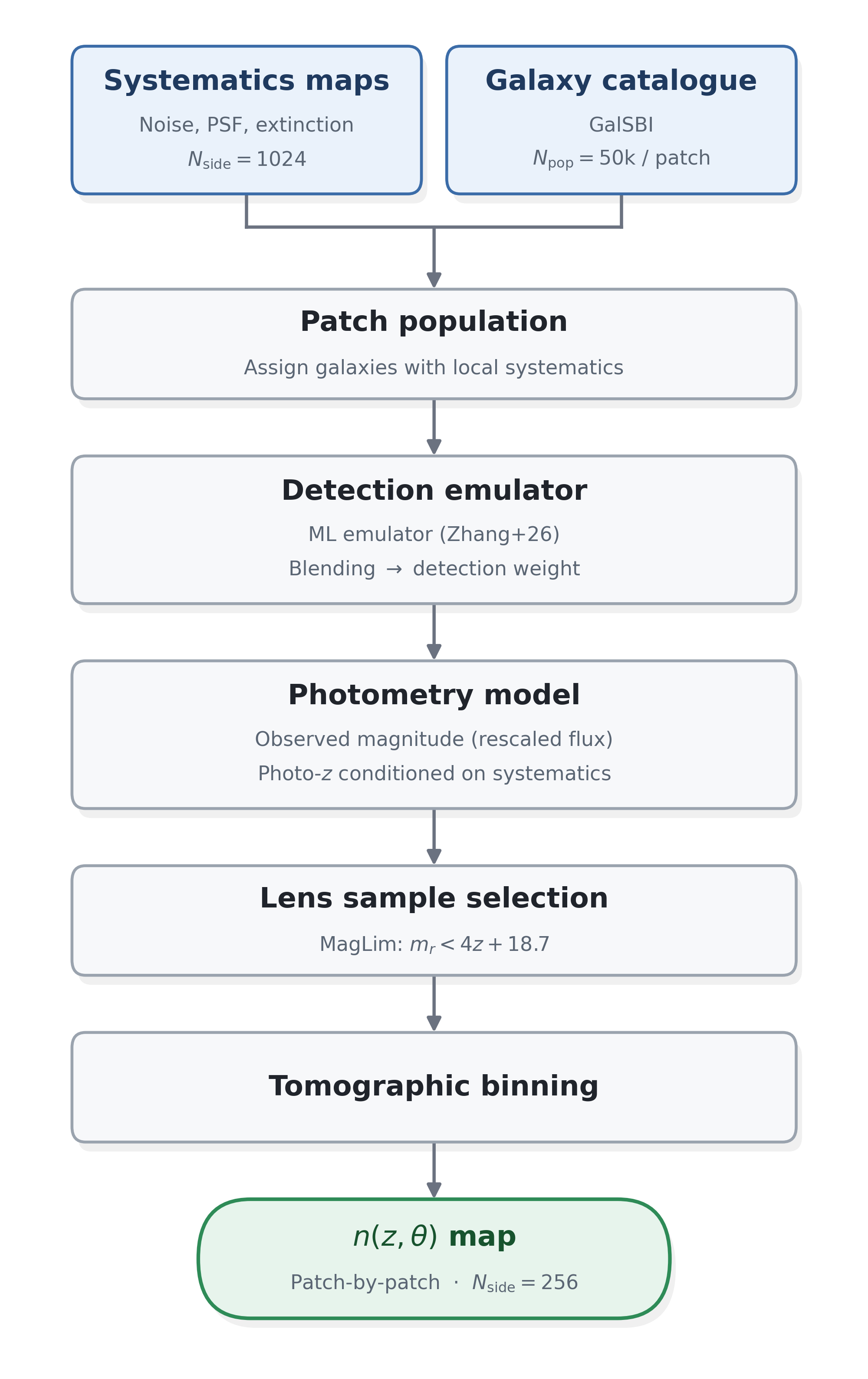}
      \caption{Overview of the methodology pipeline. Systematics maps and a mock galaxy catalog are combined to populate \textsc{HEALPix} patches.
      Each galaxy is processed through a detection emulator, a photometry model,
      and a MagLim-like lens selection, then assigned to tomographic redshift bins.
      The final output is a detection-weighted $n(z,\boldsymbol{\theta})$ map at $N_{\rm side}=256$.}
      \label{fig:flowchart}
   \end{figure}

   The spurious clustering induced by varying observing conditions has been
   tackled by a range of algorithms and applied extensively to survey data. These
   all act to remove the spurious fluctuations in galaxy \emph{number density}.
   Redshift variation shares the same cause, but concerns spatial variation of the underlying redshift
   distribution beyond its overall normalization. Because different galaxy types
   respond differently to the local observational properties, the \emph{shape}
   of $n(z)$ also varies across the sky. {This is
    where density-based corrections such as organized randoms and weight
   maps break down: both act purely on galaxy number counts and carry no
   redshift information.} Capturing the additional redshift dimension would require a non-trivial
   extension of either approach. Here we instead propose an analytical correction built upon
   a forward model of the galaxy population.

   Evaluating Terms~\ref{term2} and~\ref{term3} analytically requires a
   spatially resolved map of $n(z,\boldsymbol{\theta})$. Such a map is difficult
   to obtain directly from data or from the image simulations used for redshift
   calibration, because the finite galaxy number density renders it noise-dominated on
   small scales. In addition, $n(z,\boldsymbol{\theta})$ estimated on data can be affected by cosmic variance from genuine
   large-scale structure.

   In order to avoid these issues, we estimate
   $n(z,\boldsymbol{\theta})$ from arbitrarily large samples of mock galaxy realizations conditioned on
   local values of observational systematics, rather than directly from data. Instead of prohibitively computationally expensive image simulations, we use an emulator trained on a limited volume of such simulations to this end.
   We begin from spatial maps of observational systematics divided into small
   \textsc{HEALPix} patches and compute the mean systematic value in each patch.
   Each patch is then populated with random galaxies drawn from a mock catalog
   whose intrinsic properties, including true redshift, are known. The number of
   input galaxies can be made arbitrarily large, such that shot noise is negligible.
   Galaxies are assigned {uniform} random positions within each patch.
   By construction, no large-scale clustering is present.

   The full pipeline is illustrated in Fig.~\ref{fig:flowchart}. Systematic
   maps of pixel noise, point-spread-function full-width at half-maximum
   (PSF FWHM), and extinction are generated at
   \textsc{HEALPix} resolution $N_{\rm side}=1024$, as shown in
   Fig.~\ref{fig:sys_maps}. Each patch is populated with
   $N_{\rm pop}=50{,}000$ input galaxies. A detection emulator assigns each a
   detection probability, used as its detection weight
   (Sect.~\ref{detection}). Photometric redshift $\hat{z}$ and observed
   $r$-band magnitude $\hat{m}$ are then computed, and lens samples are
   selected on the basis of these quantities (Sect.~\ref{photo}). Galaxies
   passing the selection are assigned to six tomographic bins defined by
   $\hat{z}$, with boundaries $[0.20, 0.40, 0.55, 0.70, 0.85, 0.95, 1.05]$. The
   global redshift distribution in each bin is computed as the detection-weighted
   distribution of true redshifts, and the local distribution
   $n(z,\boldsymbol{\theta})$ is estimated patch by patch at $N_{\rm side}=256$
   (corresponding to a patch of {$\sim$0.05\,deg$^2$, or $\sim$14\,arcmin on a side}), a coarser resolution than the systematics
   maps. The remainder of this
   section describes each pipeline component in turn.

   \citet{ZhangBlending2026} developed a catalog-level detection emulator
   for Stage-III-like data, trained across a range of observing conditions to
   enable detection predictions for arbitrary conditions. They found that
   prediction improves when information about an object's surroundings is
   included, specifically the properties of the closest neighboring galaxy.
   We find this emulator particularly
   useful because it is sensitive to both the local blending environment and observing conditions.
   The input galaxies are classified in terms of detectability using this emulator.
   More details are provided in Sect.~\ref{detection}.

   Lens sample selection commonly depends on measured redshift as well as other
   quantities such as size, magnitude, and signal-to-noise ratio (SNR). In
   photometric surveys, redshifts are estimated from broadband fluxes, which is
   especially important for tomographic analyses where galaxies are assigned to
   adjacent redshift bins. We adopt a parametric model for the error in photometric redshift
   and flux, conditioned on observational systematics. Details are given in
   Sect.~\ref{photo}.

   Evaluating the impact of spatially varying $n(z,\boldsymbol{\theta})$
   further requires realistic systematic maps. We model both observational
   conditions (e.g.\ seeing and pixel noise) and observation-independent
   contaminants such as Galactic extinction, with parametric models detailed in
   Sect.~\ref{systematics}. Our aim is not to reproduce any particular survey,
   but to model a realistic level of variation in the observing conditions and
   quantify the resulting systematic effects on the cosmological analysis.

   The full code of this pipeline, dubbed \textsc{Skyvar}, is publicly available.\footnote{\url{https://github.com/zhangzzk/skyvar}}

   \subsection{Systematics}
   \label{systematics}

   Sky surveys differ in how they tile and revisit their footprints, and
   this determines the spatial structure of their systematics. In
   single-epoch surveys such as KiDS, the footprint is partitioned into
   subregions (tiles), in that case of approximately one square degree, each observed
   in a sequence of exposures at a fixed pointing with small dithers to
   mitigate bad-pixel and chip-gap artifacts. The exposures are then co-added to
   reach the survey depth. Because the exposures within a tile are acquired
   close together in time, atmospheric and instrumental conditions are nearly
   constant within a tile, while different tiles, observed days to years apart,
   sample largely independent conditions. The resulting systematics are therefore
   piecewise constant on the tile grid, with focal-plane effects such as
   pixel-to-pixel sensitivity variations adding intra-tile structure on top.

   Multi-visit surveys {reduce} this tile-level pattern. DES covers its
   footprint {ten times} per filter over six years, and LSST will
   accumulate hundreds of visits per pointing over ten years. With many visits
   spread across the survey lifetime under independently varying conditions,
   the conditions imprinted on a given sky position partially average out
   rather than tracing a single epoch. The relevant systematics map is the
   per-position depth-weighted mean over visits, with both the variance and
   the spatial coherence of the conditions reduced relative to a single-epoch
   survey. In the limit where $N$ visits on different nights are co-added for a tile, and where observational systematics are {entirely determined} by factors independent between these visits, this strategy would reduce the deviations of any tile's properties from the mean by a factor of {$1/\sqrt{N}$}. In reality, some observing conditions, especially those related to airmass and galactic latitude, are highly correlated between visits. Still, the bias amplitudes predicted from a KiDS-like observing strategy should be regarded as an upper bound for multi-visit surveys at the same
   tile size. In this work we focus on the single-epoch, KiDS-like regime,
   leaving a quantitative treatment of multi-visit averaging for future work.
   {An approach similar to the one developed here could also be applied to
   multi-visit data.}

      \begin{figure*}
      \centering
      \includegraphics[width=1\linewidth]{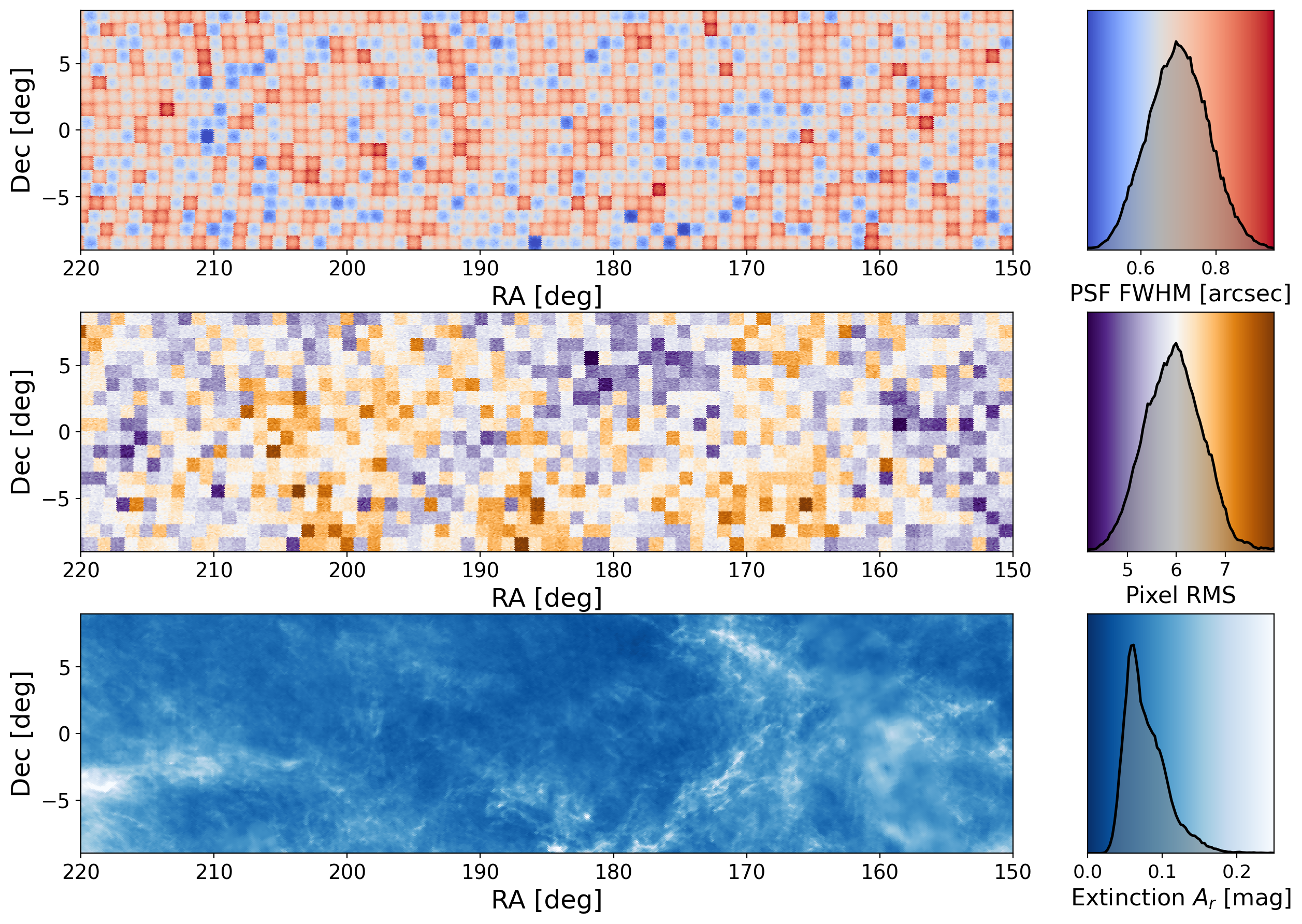}
      \caption{Simulated maps of the three observational systematics used in this
      work: $r$-band PSF FWHM ({top}), pixel RMS noise (center), and $r$-band Galactic
      extinction $A_r$ (bottom). Maps are generated at \textsc{HEALPix} resolution
      $N_{\rm side}=1024$ over a $70\,\mathrm{deg}\times18\,\mathrm{deg}$ rectangular
      footprint, with model parameters calibrated to {KiDS-Legacy} data
      \citep{liKiDSLegacyCalibrationUnifying2023,WrightKiDSLegacyData2024,YanClustering2025}. The tile structure
      is visible as piecewise variation in the noise and PSF maps, with correlated
      large-scale gradients on scales of $\sim$6 degrees and a radial intra-tile
      pattern in the PSF.}
      \label{fig:sys_maps}
   \end{figure*}

   In this work, we model KiDS-like systematics using square tiles with
   systematics varying both between and within tiles, with global distributions
   calibrated against the {KiDS-Legacy} data
   \citep{liKiDSLegacyCalibrationUnifying2023, YanClustering2025}. {The simulated maps are shown in Fig. \ref{fig:sys_maps}.} The scale of
   systematic variation depends strongly on the tile size of the specific survey:
   KiDS tiles cover approximately 1~deg$^{2}$, while the fields of view of DES
   and LSST are approximately {4~deg$^{2}$ and 9~deg$^{2}$}, respectively.
   Although our default setup is KiDS-like, we explore the impact of tile size in
   Appendix~\ref{sec:impact_tile_size}.

   Following \citet{YanClustering2025} and the associated code
   repository\footnote{\url{https://github.com/yanzastro/tiaogeng}}, we model
   observational systematics on a \textsc{HEALPix} grid within a rectangular
   survey footprint spanning $150^{\circ} \leq \alpha \leq 220^{\circ}$ in
   right ascension and $-9^{\circ} \leq \delta \leq 9^{\circ}$ in declination
   (Table~\ref{tab:model_params}). The modeled systematics are $r$-band PSF FWHM,
   pixel root-mean-square (RMS) noise, and $r$-band Galactic extinction. Beyond these, we note that other
   spatially varying quantities can also affect selection, including sky brightness
   from non-Galactic foregrounds, stellar density, and bright-star masking.
   Atmospheric and Galactic extinction in multiple bands can also affect the
   observed colors, which are critical for photo-$z$ estimation and tomographic
   binning.

   The noise map is tile-wise constant at leading order, with correlated inter-tile
   variation in the tile means and additional pixel-scale jitter. The seeing contains
   both correlated inter-tile variation and intra-tile structure. For
   PSF and noise, tile-level means are drawn from a Gaussian random field on
   tile centers with a radial basis function covariance,
   \begin{equation}
      \mathbf{C}_{ij}=\sigma_{\mathrm{corr}}^{2}\exp\!\left(-\frac{D_{ij}^{2}}{2\ell_{\mathrm{corr}}^{2}}
      \right),
   \end{equation}
   where $D_{ij}^{2}$ is the squared angular distance between tile centers $i$ and
   $j$. An independent tile scatter $\mathcal{N}(0,\sigma_{\mathrm{uncorr}})$ is
   then added.

   The noise map is piecewise constant per tile plus pixel jitter:
   \begin{equation}
      N(\mathbf{\theta}) = n_{t}+ \eta_{\mathrm{pix}}, \qquad \eta_{\mathrm{pix}}
      \sim\mathcal{N}(0,\sigma_{\mathrm{pix}}).
   \end{equation}
   {Here $n_{t}$ is the tile-level mean drawn from the Gaussian random field
   above, and $\sigma_{\mathrm{pix}}$ sets the amplitude of the pixel-level jitter.}

   {The PSF varies within a tile due to focal plane dependence and spatial atmospheric structure, and across observations due to varying observing conditions. We model the intra-tile PSF structure using an isotropic Gaussian profile centered on each tile,}
   \begin{equation}
      G(\mathbf{\theta})=\exp\!\left[-\frac{x^{2}+y^{2}}{2s_{\mathrm{intra}}^{2}}
      \right],
   \end{equation}
   with local coordinates $(x,y)$ in degrees {and $s_{\mathrm{intra}}$ the intra-tile Gaussian scale}. {Its amplitude varies across tiles to mimic the inter-tile structure.} The implemented model is
   \begin{equation}
      \mathrm{PSF}(\mathbf{\theta}) = s_{t}\,[2-G(\mathbf{\theta})] + \epsilon_{\mathrm{pix}}
      , \qquad \epsilon_{\mathrm{pix}}\sim\mathcal{N}(0,\sigma_{\mathrm{pix}}){,}
   \end{equation}
   {where $s_{t}$ is the tile-level PSF value (again drawn from the Gaussian
   random field) and the factor $2-G$ imprints the radial intra-tile pattern.}

   {We include Galactic extinction as a representative observation-independent
   systematic for completeness, though it is subdominant to the PSF and noise
   variations.} Galactic extinction is evaluated from the SFD dust map \citep{SFD1998,GreenDustmaps2018}
   as
   \begin{equation}
      A_{r}= 2.285\,E(B-V).
   \end{equation}
   $E(B-V)$ is sampled from the SFD map at the selected simulated footprint coordinates,
   which lie close to the KiDS northern patch. We note that, in a real analysis, the
   dereddening applied to the photometry is itself imperfect: because the
   reddening in a given band depends on the source spectral energy distribution,
   a single $A_r$ per pixel cannot fully correct a heterogeneous galaxy
   population and leaves a residual, spatially varying color bias.

   The goal of this work is to quantify the effect at a realistic level rather
   than to provide a {calibration}, which would require using actual
   systematic maps from data. Here, we use parametric models whose free parameters
   are tuned to match the {KiDS-Legacy} data
   \citep{liKiDSLegacyCalibrationUnifying2023,WrightKiDSLegacyData2024,YanClustering2025}.
   Specifically, the global mean and
   standard deviation of the noise and PSF FWHM are set to match the {KiDS-Legacy}
   distributions. The uncorrelated tile scatter of both quantities is the dominant
   source of variation, while the correlated component adds secondary large-scale
   correlations, with correlation lengths chosen to approximately match those
   observed in {KiDS-Legacy} systematics maps \citep{YanClustering2025}. Other
   parameters are of secondary importance: the pixel-level jitter of both noise and
   PSF FWHM is set to negligible levels, and the intra-tile Gaussian scale is set
   to the order of the tile size. All adopted parameter values are summarized in
   Table~\ref{tab:model_params}.

   \begin{table}
      \caption{Parameters used for the mock systematics maps and photo-$z$ model.}
      \label{tab:model_params}
      \centering
      \begin{tabular}{lll}
         \hline
         \hline
         Parameter                               & Value                      & Description                             \\
         \hline
         \multicolumn{3}{l}{Footprint and tiling} \\
         $\alpha_{\min}, \alpha_{\max}$          & $150^{\circ}, 220^{\circ}$ & RA range                                \\
         $\delta_{\min}, \delta_{\max}$          & $-9^{\circ}, 9^{\circ}$  & Dec range                               \\
         $\mathrm{tile\_size}$                   & $1.0^{\circ}$              & Tile side length                        \\
         \hline
         \multicolumn{3}{l}{Noise model}          \\
         $\mu_{0}^{\rm noise}$                   & $6.0$                      & Global mean pixel noise                 \\
         $\sigma_{\rm corr}^{\rm noise}$         & $0.5$                      & Correlated tile-level scatter           \\
         $\ell_{\rm corr}^{\rm noise}$           & $6.0^{\circ}$              & Correlation length                      \\
         $\sigma_{\rm uncorr}^{\rm noise}$       & $0.6$                      & Uncorrelated tile scatter               \\
         $\sigma_{\rm pix}^{\rm noise}$          & $0.02$                     & Pixel-level jitter                      \\
         \hline
         \multicolumn{3}{l}{PSF model}            \\
         $\mu_{0}^{\rm psf}$                     & $0.7$                      & Baseline seeing                         \\
         $\sigma_{\rm corr}^{\rm psf}$           & $0.03$                     & Correlated tile-level scatter           \\
         $\ell_{\rm corr}^{\rm psf}$             & $6.0^{\circ}$              & Correlation length                      \\
         $\sigma_{\rm uncorr}^{\rm psf}$         & $0.08$                     & Uncorrelated tile scatter               \\
         $s_{\rm intra}$                         & $1.0^{\circ}$              & Intra-tile Gaussian scale               \\
         $\sigma_{\rm pix}^{\rm psf}$            & $0.02$                     & Pixel-level jitter                      \\
         \hline
         \multicolumn{3}{l}{Photo-$z$ model}      \\
         {$\sigma_{\rm phot}$}                      & $0.0376$                   & Base photo-$z$ scatter at $m_{\rm ref}$ \\
         $\sigma_{\rm int}$                      & $0.0212$                   & Intrinsic photo-$z$ floor               \\
         $m_{\rm ref}$                           & $21$                       & Reference magnitude                     \\
         $\mathrm{RMS}_{\rm ref}$                & $6.0$                      & Reference noise level                   \\
         $\mathrm{PSF}_{\rm ref}$                & $0.7$                      & Reference PSF FWHM                      \\
         $\mathrm{maglim}_{0}$                   & $4$                        & MagLim slope                            \\
         $\mathrm{maglim}_{1}$                   & $18.7$                     & MagLim intercept                        \\
         $\mathrm{maglim}_{2}$                   & $17$                       & Bright-end MagLim cut                   \\
         \hline
         \hline
      \end{tabular}
   \end{table}

   \subsection{Galaxy catalog}
   \label{catalog}

   {We use GalSBI to generate a mock galaxy catalog \citep{FischbacherGalSBI2025}.
   GalSBI is a phenomenological galaxy population model. It adopts an analytic
   parametrization of galaxy luminosity functions, morphologies, and SEDs.
   The model parameters are constrained via simulation-based inference
   against HSC deep-field imaging. We follow the fiducial pipeline and
   posterior choice from \url{https://cosmo-docs.phys.ethz.ch/galsbi/}.

   We start from a GalSBI-generated mock catalog, using it as
   an empirical pool of intrinsic galaxy properties rather than
   a fixed sky realization. The input catalog contains
   $4,337,172$ galaxies with intrinsic redshift, $r$-band magnitude,
   size, Sérsic index, and ellipticity. We apply the same quality cuts
   used when training the selection emulator \citep{ZhangBlending2026},
   leaving $4,018,873$ galaxies in the source pool.

   For each $\textsc{HEALPix}$ simulation pixel at $N_{\rm side}=1024$, we draw
   $N_{\rm pop}$ galaxies independently and uniformly from this filtered pool,
   with replacement. These draws are then assigned the observing conditions of
   that pixel: the local PSF FWHM, pixel noise, and Galactic extinction.  The
   resulting per-pixel catalogs are passed through the detection and
   post-selection pipeline to construct the tomographic samples.}

   \subsection{Detection emulator}
   \label{detection}

   Only a small fraction of the galaxies present in a given patch of sky
   are recovered by a survey, because we observe a noisy, pixellated,
   PSF-convolved image rather than the underlying true scene. Detection
   algorithms in real surveys typically flag concentrated flux above a predefined
   threshold, so the detection efficiency depends on both the intrinsic
   properties of a galaxy and the local observing conditions.

   The detection process is conventionally studied using image simulations, in
   which the algorithm is validated through precise forward modeling of the
   galaxy population and observational setup, an approach previously used to
   characterize survey selection in cosmological inference
   \citep[e.g.,][]{FischbacherGalSBI2025}. However, due to the high
   computational cost of such simulations, emulation is often used as an
   efficient alternative.
   \citet{ZhangBlending2026} developed a machine-learning-based detection emulator
   trained on KiDS-like image simulations, in which the true properties of input
   galaxies are known. The emulator learns the mapping between these properties and
   the detection outcome.

   As noted above, the detection of an object depends not only on its own intrinsic
   properties but also on nearby objects in the focal plane, i.e., blending. For
   example, an object with a very bright nearby neighbor is less likely to be
   detected than when isolated. \citet{ZhangBlending2026} therefore built the
   emulator to take the blending configuration as additional input features: the
   magnitude, size, and axis ratio of the closest neighbor within 3\,arcsec, as
   well as the angular separation between the neighbor and the primary galaxy. The
   emulator then outputs the detection probability of the primary galaxy.

   The model also incorporates the local observing conditions. Following
   Eqs.~16--18 of \citet{ZhangBlending2026}, the input features are rescaled as
   \begin{equation}
      \label{eq:scale1}\tilde{r}_{e}= \frac{r_{e}}{\sqrt{r_{\rm PSF}^{2}+r_{e}^{2}}}
      ,
   \end{equation}

   \begin{equation}
      \label{eq:scale2}\tilde{\theta}= \frac{\theta}{\tilde{r}_{e,\rm primary}},
   \end{equation}

   \begin{equation}
      \label{eq:scale3}\tilde{f}= \frac{f}{\mathrm{RMS}\cdot\pi(r_{\rm PSF}/r_{\rm pixel})^{2}}
      ,
   \end{equation}
   where $r_{e}$ is the effective radius, $r_{\rm PSF}$ is the PSF half-light radius {(the PSF FWHM of Sect.~\ref{systematics} converted to a half-light radius)},
   $\theta$ is the angular separation between the primary galaxy and its closest
   neighbor, $f$ is the flux of the neighbor, $\mathrm{RMS}$ is the pixel noise,
   and $r_{\rm pixel}$ is the pixel scale.

   This rescaling is motivated by the fact that detection is governed by
   apparent photometry: the PSF-convolved size of the galaxy and its
   flux-to-noise ratio effectively determine the detection outcome. \citet{ZhangBlending2026} validated
   this rescaling by demonstrating that the prediction accuracy of the pre-trained
   emulator is maintained on test sets with varying observing conditions. {The emulator lets us predict detection probabilities across the observing
   conditions in our mock at low computational cost. Its inputs are each galaxy's
   {true} flux and size, the properties of its nearest neighbor, and the local
   observing conditions, combined through the rescalings above. The mapping from these inputs to detection
   probability is learned from image simulations with known input catalogs.}

   \subsection{Photometry}
   \label{photo}

   {The galaxies used
   for clustering measurements --- referred to as lens galaxies in $2\times2$pt
   analyses combining clustering and weak lensing --- must balance two competing
   requirements: a larger sample reduces shot noise, while a narrow and
   accurately characterized redshift kernel is needed to extract clustering
   information cleanly, since the clustering signal is sensitive to the width
   of $n(z)$.} Lens-sample selection is therefore a trade-off between sample size
   and photo-$z$ uncertainty (e.g., \citealt{2020TanoglidisLens}), and lens galaxies are
   subject to more aggressive selection criteria than the source galaxies used
   for shear measurements, such as tighter magnitude limits.

   A common choice is the MagLim selection, a redshift-dependent magnitude
   cut, e.g., of the general form
   \begin{equation}
      \hat{m}_{\rm band} < a\,\hat{z} + b,
   \end{equation}
   where $\hat{m}_{\rm band}$ is the observed magnitude in a chosen band
   and $\hat{z}$ is the photometric redshift. The slope $a$ and intercept $b$
   are tuned to optimize the cosmological information per unit shot noise
   \citep{PorredonDESY3MagLim2021}. The DES Y3 and Y6 analyses adopt $a=4$
   and $b=18$ in the $i$-band. For comparison, redMaGiC
   \citep{RozoRedMaGiC2016} (used in DES Y1) applies a stricter
   color-based selection, keeping only red luminous galaxies whose photo-$z$
   uncertainty is minimal, at the cost of a smaller sample.

   In this work we keep the DES slope $a=4$ but adapt the band and the
   intercept. We work in the $r$-band only. Notably, DES and KiDS-Legacy have substantially deeper \(r\)-band than \(i\)-band imaging, with the difference in their depths being approximately {$0.6$ and $0.9$ mag, respectively} {\citep{WrightKiDSLegacyData2024,DESdata2026}}. {Red, high-redshift galaxies are
   often brighter in redder bands, so an $i$- or $z$-band cut could be preferable at
   the highest redshifts; we nonetheless adopt the $r$-band for its greater depth in
   the surveys considered here.} Our default lens selection is
   therefore
   \begin{equation}
      \hat{m}_{r}< 4\,\hat{z}+ 18.7.
   \end{equation}
   We also investigate different choices of the intercept
   in {Sect.} \ref{sec:impact_sample_selection}.

   We require a model for the observed magnitude and photometric redshift.
   We model the observed $r$-band magnitude by adding Gaussian photometric
   noise to the input magnitude,
   \begin{equation}
      \hat{m}_{r}=m_{r}+\epsilon_{m},\qquad
      \epsilon_{m}\sim\mathcal{N}(0,\sigma_{m}),\qquad
      \sigma_{m}=\frac{2.5}{\ln 10}\frac{1}{\mathrm{SNR}}.
      \label{eq:mag_error}
   \end{equation}
   Here SNR is the per-galaxy signal-to-noise ratio computed from its flux,
   effective PSF-convolved size, and the local pixel noise{,}
   \begin{equation}
       {{\rm SNR} = \frac{f}{\mathrm{RMS}*\sqrt{4\pi(r_{\rm PSF}^2+r_e^2)/r_{\rm pixel}^2}}.}
   \end{equation}
   {Photometric redshifts rely on multiple-band magnitudes, whose errors may scale
   differently with observing conditions. To simplify the problem, we assume
   that they change coherently and ignore the color dependence of conditions,
   thus modeling the photo-$z$ uncertainty from $r$-band
   photometry alone.}

   The photometric redshift is modeled as a Gaussian distribution around the
   truth,
   \begin{equation}
      \hat{z} {\sim} \mathcal{N}(z_{\rm true}, \sigma_{z})\;,
   \end{equation}
   where $\sigma_{z}$ is the photo-$z$ uncertainty. We assume that this
   uncertainty can be decomposed into an intrinsic component and a component resulting from photometric noise,
   \begin{equation}
      \sigma_{z}= \sqrt{\sigma_{\rm int}^{2}+ \sigma_{\rm phot}^{2}}.
   \end{equation}
   The intrinsic component arises because redshifts are estimated from the
   integrated flux in a given set of discrete photometric bands: {degeneracies between
   redshift and spectral type, together with intrinsic color scatter among
   galaxies of the same broad spectral type, cause uncertain redshift estimates
   even with perfect photometry.} The
   photometric component arises from noisy flux measurements. {We model it
   as inversely proportional to the rescaled flux $\tilde f$ of
   Eq.~\ref{eq:scale3}, evaluated using the flux of the galaxy whose redshift is
   being estimated:}
   \begin{equation}
      \sigma_{\rm phot}\propto {\tilde f^{-1}}.
      \label{eq:sigma_phot_prop}
   \end{equation}
   To fix the normalization of $\sigma_{z}$, in particular the proportionality coefficient in Eq.~\ref{eq:sigma_phot_prop} and the value of $\sigma_{\rm int}$, we anchor it to two reference
   samples drawn from \citet{JamieDC3R22024}. Here the authors evaluated the photo-$z$
   uncertainty on matched galaxy colors in both COSMOS and the DESI-KV sample
   (a DESI cross-match against KiDS--VIKING that is representative of
   {KiDS-Legacy} observing conditions). We approximate the COSMOS
   median photo-$z$ uncertainty as the intrinsic floor $\sigma_{\rm int}$,
   since COSMOS is a deep survey with high-quality multi-band photometry and
   its residual uncertainty is dominated by spectral-type-redshift
   degeneracies rather than measurement noise. The exact value is tied to
   {the filters used}. The COSMOS photometry here takes $u$ band imaging from CFHT,
   $griz$ from Subaru Suprime Cam,
   and YJH from the {UltraVISTA} survey \citep{Masters2015}. The quadrature difference
   between the DESI-KV and COSMOS medians gives the photometric component
   $\sigma_{\rm phot}$ at the DESI-KV typical observing conditions. This
   value, together with the mean DESI-KV $r$-band magnitude, defines the
   reference point at which the proportionality
   $\sigma_{\rm phot}\propto\tilde{f}^{-1}$ is normalized. Numerically, this
   procedure gives $\sigma_{\rm int} = 0.0212$ (COSMOS median) and
   $\sigma_{\rm phot} = 0.0376$ at the reference point
   $(m_{\rm ref}, \mathrm{RMS}_{\rm ref}, \mathrm{PSF}_{\rm ref})=(21,\,6.0,\,0.7)$,
   from which $\sigma_{\rm phot}$ at any other galaxy magnitude and observing
   condition follows from Eqs.~\ref{eq:scale3} and~\ref{eq:sigma_phot_prop}.
   The full set of model parameters is summarized in
   Table~\ref{tab:model_params}.

   \section{Redshift variation}

   \label{sec:main_results}
   In this section we quantify how a spatially varying $n(z,\boldsymbol{\theta})$
   alters the angular clustering signal $w(\theta)$. {We present the main effect here.}
   {The impact} of tile size (Appendix~\ref{sec:impact_tile_size}),
   lens-sample selection (Appendix~\ref{sec:impact_sample_selection}), and the choice
   of definition for the global $\bar{n}(z)$ (Appendix~\ref{sec:impact_n_def}) is
   assessed in appendices.

   Throughout this section we set the linear galaxy bias to
   $\bar{b}(z)=1$. In reality $\bar{b}(z)$ evolves with redshift, and we adopt
   a more realistic redshift-dependent bias in Sect.~\ref{sec:cosmo} when
   translating the effect to cosmological-parameter shifts. {Our main quantity is the ratio $w_{\rm true}/w_{\rm model}$.
   A bias amplitude that is constant within a tomographic bin cancels exactly
   between numerator and denominator, so the ratio is insensitive to that
   amplitude. Redshift evolution of the bias within a bin changes the weighting
   of the redshift integrals and does not, in general, cancel exactly.}

   \subsection{Spread in the redshift-kernel width}
   \label{sec:nz_variation}

\begin{figure*}
      \centering
      \includegraphics[width=1\linewidth]{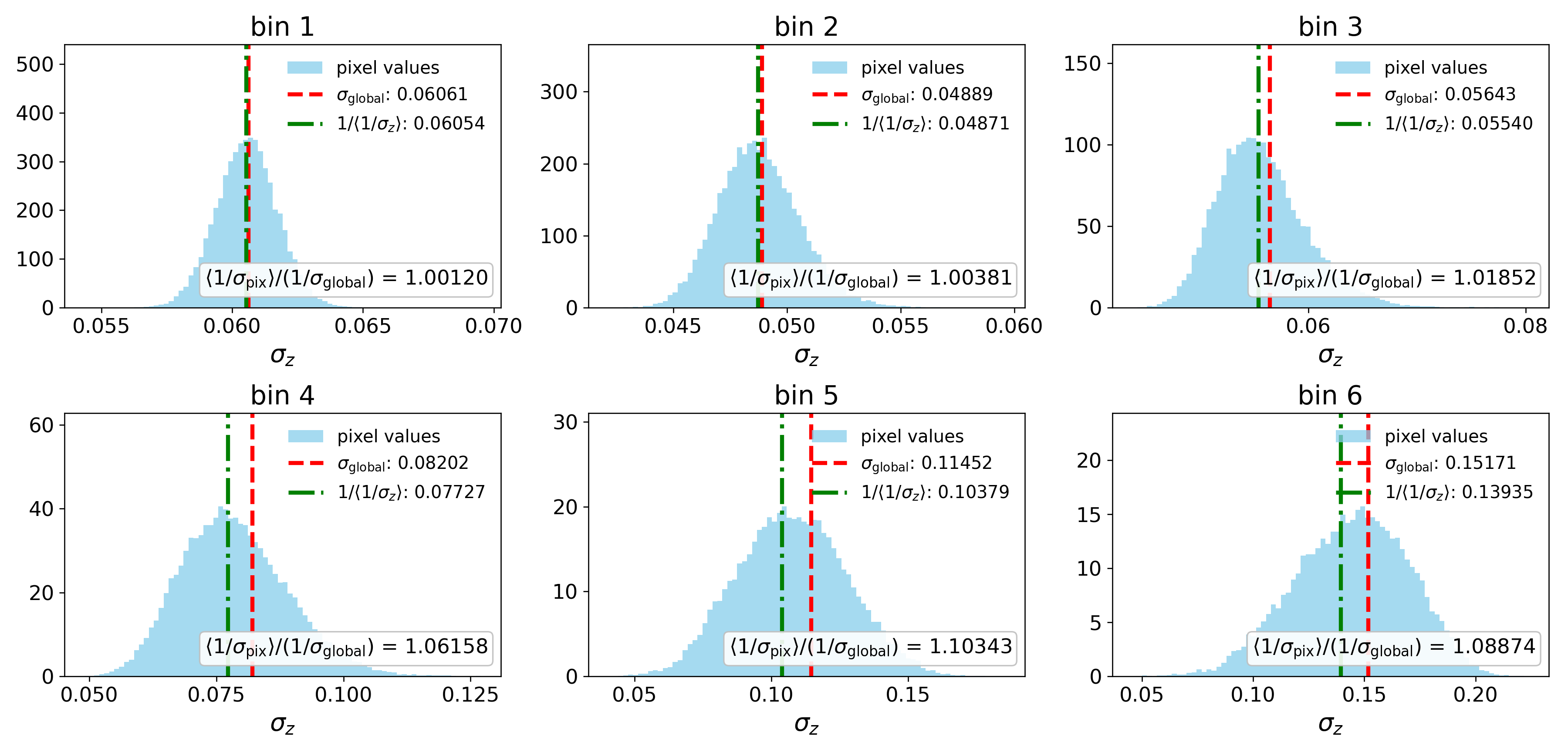}
      \caption{Pixel-to-pixel variation of the redshift-kernel width across
      the six tomographic bins. Each panel shows the
      distribution of the local width $\sigma_{\rm pix}$ (the standard deviation $\sigma_z$ of
      $n(z,\boldsymbol{\theta})$) over \textsc{HEALPix} pixels at $N_{\rm side}=256$;
      the red dashed line marks the width $\sigma_{\rm global}$ of the global
      distribution $\bar{n}(z)$, and the green dash-dotted line the
      harmonic-mean width $1/\langle 1/\sigma_{\rm pix}\rangle$ that sets the
      pixel-averaged small-scale clustering amplitude. The spread broadens toward higher redshift, where
      selection and photo-$z$ effects are strongest. Because the small-scale
      clustering amplitude scales as $w(\theta=0)\propto 1/\sigma_z$, the inverse-width
      ratio $\langle 1/\sigma_{\rm pix}\rangle/(1/\sigma_{\rm global})$ quoted in each
      panel approximates the {enhancement factor for} the auto-correlation amplitude; it
      grows from $\approx 1$ in the lowest bins to $\sim 1.09$ in the highest,
      mirroring the per-bin bias in the tomographic $w(\theta)$
      (Fig.~\ref{fig:tomographic_bins_w}).}
      \label{fig:nz}
   \end{figure*}

   We begin with the redshift distribution itself. Spatially varying observing
   conditions imprint a pixel-to-pixel spread in the mean of the redshift distribution. {The global redshift variance is the sum of the mean within-pixel
   variance and the variance of the pixel means, using the same pixel weights
   as in the global distribution. Thus, the mean within-pixel variance cannot
   exceed the global variance, although an individual pixel's variance can.}
   Fig.~\ref{fig:nz} shows the distribution of the local width $\sigma_{\rm pix}$
   across \textsc{HEALPix} pixels for each tomographic bin. The spread increases toward
   higher redshift, where selection and photo-$z$ effects are strongest (as we detail
   below). The standard deviation of the redshift distribution directly sets the small-scale clustering
   amplitude: for a Gaussian $n(z)$ of standard deviation $\sigma_{z}$, the amplitude
   scales as
   \begin{equation}
      w(\theta=0) \;\propto\; \int \mathrm{d}z\, n^{2}(z)
                  \;\approx\; \frac{1}{\sigma_{z}},
      \label{eq:w_scaling}
   \end{equation}
   so a narrower $n(z)$ produces stronger apparent clustering. Averaged over
   tiles, evaluating the amplitude from the global $\bar{n}(z)$ rather than tile by
   tile misestimates the clustering amplitude by the inverse-width ratio
   $\langle 1/\sigma_{\rm pix}\rangle/(1/\sigma_{\rm global})$, which gives a simple
   analytic estimate of the {enhancement factor for} the auto-correlation amplitude. The
   resulting per-bin factors grow from $\approx 1.00$ in the lowest redshift bins to
   $\sim 1.09$ in the highest redshift bins. {This provides an expectation of the clustering enhancement that the full
   computation below reproduces in detail.}

   \subsection{Noise validation}
   \label{sec:analytic_estimate}

   Before turning to the full computation, we verify that the pixel-to-pixel
   spread just shown reflects genuine selection effects rather than sample variance.
   One caveat of drawing galaxies
   independently in each tile is that {this} introduces Poisson fluctuations in
   $n(z,\boldsymbol{\theta})$ from tile to tile. {However, {these fluctuations are}} expected to be subdominant for
   large $N_{\rm pop}$, which we confirm with a simple estimate. The inverse-width
   ratio of Eq.~\ref{eq:w_scaling}, $\langle 1/\sigma_{\rm pix}\rangle/(1/\sigma_{\rm
   global})$ (quoted per bin in Fig.~\ref{fig:nz}), serves as {an approximate estimate}
   of the enhancement {of} $w(\theta)$ due to varying $n(z)$ across tiles.

   {A non-unit ratio does not by itself establish a selection-induced clustering
   enhancement. Finite samples drawn independently from the same redshift distribution
   produce fluctuations in the local kernels and can bias the inverse-width ratio
   away from unity. These fluctuations have zero expected cross-correlation between
   distinct pixels. At separations within a pixel, however, assigning one estimated
   kernel to the whole pixel can produce a finite-sampling contribution. We use a
   simplified numerical experiment to estimate the sampling contribution to the
   inverse-width ratio.}
   {We draw independent samples from a toy Gaussian redshift kernel to
   represent the pixels. We calculate the inverse-width ratio using the standard
   deviation of each sample and that of the pooled samples, analogous to the
   local and global $\sigma_z$.}
   {The sample size is set to the mean number of selected galaxies in a pixel,}
   833, 438, 256, 219, 210 and 324 respectively for the six tomographic bins.
   {We obtain} 1.0015, 1.0029, 1.0050, 1.0057, 1.0060 and 1.0039, which are subdominant
   relative to the factors estimated from data, particularly for the
   higher-$z$ bins.

   {Using this ratio to estimate the small-scale enhancement of $w(\theta)$
   assumes that $n(z,\boldsymbol{\theta})$ is constant within each pixel.} This is reasonable for $N_{\rm side}=256$, given the typical
   variation scale of systematics is much larger.

   \subsection{Enhancement of the angular clustering}
   \label{sec:w_enhancement}

\begin{figure*}
      \centering
      \includegraphics[width=1\linewidth]{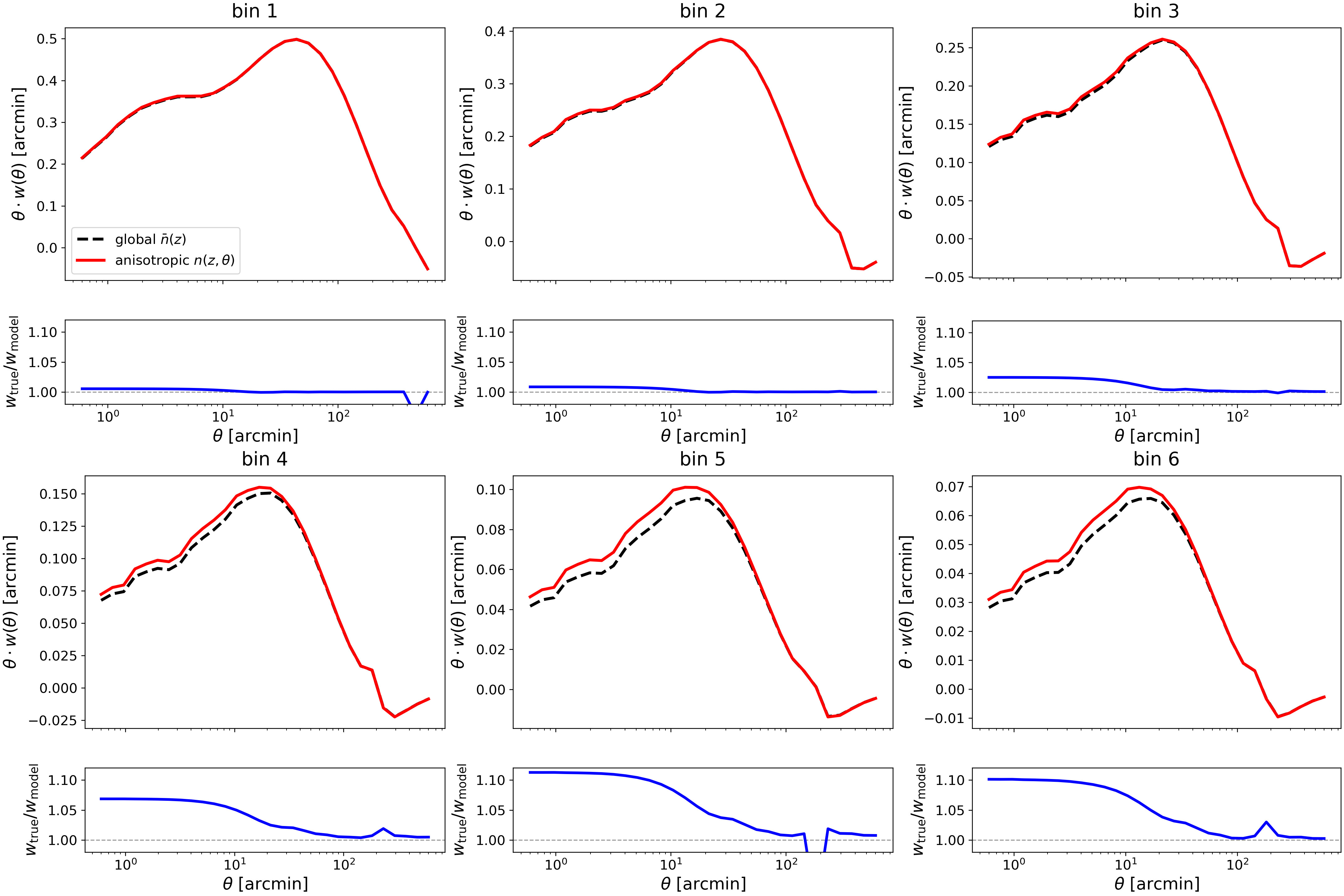}
      \caption{{Angular clustering shown as $\theta w(\theta)$ in the upper panels} for each of the
      six tomographic bins (panels, increasing redshift {from top left to bottom right}). {Red}
      lines show $w_{\rm true}$, computed using the spatially varying redshift
      distribution $n(z,\boldsymbol{\theta})$ via Eqs.~\ref{term1}--\ref{term3}. {Black lines show} $w_{\rm model}$, computed assuming a spatially constant
      $\bar{n}(z)$ (Eq.~\ref{term1} only). The ratio $w_{\rm true}/w_{\rm model}$
      is shown in the lower sub-panels. Modeling with a constant $\bar{n}(z)$ underestimates $w(\theta)$ at small
      angular scales, with the bias growing for higher-redshift bins.
      {Features in the ratios at separations of several hundred arcminutes
      occur near zero crossings of $w(\theta)$, where the ratios are numerically unstable.}
      The same applies to similar figures appearing in {the Appendix}.}
      \label{fig:tomographic_bins_w}
   \end{figure*}

   Fig.~\ref{fig:tomographic_bins_w} presents the main result of this work on the data vector side.
   We show $w(\theta)$, computed assuming either a spatially constant or a spatially varying
   redshift distribution, referred to as $w_{\rm model}$ and $w_{\rm true}$
   respectively. These correspond to either only Eq.~\ref{term1} or Eqs.~\ref{term1}--\ref{term3}. The underlying
   $n(z,\boldsymbol{\theta})$ maps and $\Bar{n}(z)$ are estimated following the mock pipeline described
   in Sect.~\ref{mocks}. The redshift integrals are evaluated in thin shells.
   Within each shell, the angular power spectrum of the matter overdensity is
   computed with \textsc{pyccl} {\citep{ChisariCCL2019}} for the given cosmology. Rather than performing a
   full mode-coupling (mask) deconvolution, we approximate this by dividing the
   per-shell power spectrum by the observed sky fraction $f_{\rm sky}$, i.e.\ the
   ratio of the footprint area to the full sky. The resulting angular power spectra
   are then transformed to configuration space to give $w(\theta)$.

   As shown, $w_{\rm model}$ based on a constant $\bar{n}(z)$ is in broad agreement
   with $w_{\rm true}$, but underestimates the amplitude by a few percent at angular
   scales below $\approx 100$\,arcmin. For our KiDS-like setup, modeling with a constant $\bar{n}(z)$
   {underestimates} the true $w(\theta)$ by roughly 10\% at separations below 10\,arcmin
   for the tomographic bin $(0.95, 1.05]$, dropping to $\sim$2\% for bin $(0.55,
   0.70]$. Lower redshift bins are nearly unaffected. The characteristic scale reflects the spatial
   structure of observational systematics, which are dominated by tile-level variation
   ($\sim$60\,arcmin) with secondary intra-tile contributions. At separations exceeding
   this scale, the two-point correlation becomes progressively insensitive to these
   variations and converges to the model that assumes a global redshift distribution. This is expected in the limit where the two local $n(z)$s a galaxy pair is drawn from are independent of one another and hence have a cross-variance equal to that of $\bar{n}(z)$.

   The underestimation grows with increasing redshift bin, for two main reasons.
   First, higher-redshift galaxies are {fainter in apparent magnitude} and therefore more
   susceptible to selection bias at the detection stage, since faint objects are more
   sensitive to observing conditions. Second, faint galaxies also carry larger
   photometric uncertainties: since bin assignment relies on measured magnitudes and
   photometric redshifts, the more noise-dominated photometric redshift error becomes more susceptible to systematic perturbations, and
   a galaxy near a bin boundary can scatter into either the current or the adjacent bin
   under slight changes in observing conditions.

   \section{Cosmological constraints}
   \label{sec:cosmo}

   Neglecting the effect of the spatial $n(z,\boldsymbol{\theta})$ variation
   when modeling clustering data could lead to internal tension or biases in cosmological inference. To quantify this, we
   fit a baseline model assuming a spatially constant $\bar{n}(z)$ to the contaminated
   data vectors and assess the impact on cosmological inference.

   We
   construct the contaminated data vectors by multiplying the baseline predicted
   $w(\theta)$ by the ratio $w_{\rm true}{(\theta)}/w_{\rm model}{(\theta)}$ from Sect.~\ref{sec:main_results}.
   We adopt the baseline model and inference pipeline from the DES Year~3 analysis \citep{DESY32022} using \textsc{CosmoSIS} \citep{Zuntz2015}, described in
   Sect.~\ref{sec:cosmo_setup}. We run a joint analysis of galaxy clustering and
   galaxy--galaxy lensing (GGL). The lensing data vector remains mostly
   unaffected by the redshift distribution variation {\citep{Kong2026,HangLSST2024}}
   and thus serves as an anchor. {As in a blinded cosmological analysis, we test the consistency of the
   contaminated clustering with lensing before examining the inferred parameters}
   (Sect.~\ref{ppd_analysis}). We then quantify the resulting shifts in the inferred
   parameters (Sect.~\ref{sec:cosmo_cosmo}).

   \subsection{Analysis setup}
   \label{sec:cosmo_setup}

   We perform a simplified, so-called $2\times2$pt analysis. {We adopt a
   fiducial cosmology based on the Planck 2018 TT+TE+EE+lowE+lensing posterior
   mean parameters} \citep[Table~2]{PlanckVI2020},
   {with the adopted values listed} in Table~\ref{tab:fixed-marginal-constraints}.
   The non-linear matter power spectrum is computed with HMcode2020 \citep{Mead2021} using a
   feedback temperature $\log_{10}(T_{\rm AGN/K})=7.8$. The linear galaxy bias in six tomographic lens
   bins takes fiducial values $b = \{1.5, 1.8, 1.8, 1.9, 2.3, 2.3\}$ from the MagLim sample validation of \citet[{their}~Table~2]{Krause2021}.
   Following the simplified setup of \citet[Sect.~4.2]{Emas2026}, we do not include well-studied observational systematics such as intrinsic alignments,
   {redshift-calibration uncertainties}, shear calibration bias, or magnification.

   The covariance matrix is an analytical Gaussian covariance computed with CosmoCov \citep{KrauseEifler2017, Fang2020},
   without super-sample covariance or non-Gaussian contributions. We apply scale cuts of $R_{\rm min} = 6~\mathrm{Mpc}/h$
   for galaxy-galaxy lensing $\gamma_t$ and $R_{\rm min} = 8~\mathrm{Mpc}/h$ for
   $w(\theta)$, converted to redshift-dependent angular scales per lens bin, similar
   to the strategy of \citet{Krause2021}.
   {These scale cuts are commonly used in cosmological analyses to avoid
   uncertain small-scale physics.} {They also remove part of the regime where the
   systematic effect studied here is strongest. While we apply these conservative scale cuts,
   we expect the effect to have a larger impact on analyses that indeed include nonlinear}
   scales.
   The posterior is sampled with the Nautilus nested sampler \citep{Lange2023} using 6000 live points.

   We construct three survey-like setups to assess the significance of the effect
   across different survey configurations,
   building on the footprint-level variation analysis presented above.
   The statistical power of each setup is governed primarily by the survey area and galaxy number density.
   We adopt areas of 1260~$\text{deg}^2$ (KiDS-like),
   5000~$\text{deg}^2$ (DES-like), and 12300~$\text{deg}^2$ (LSST-like). For the number density,
   we take the per-bin values reported for KiDS-1000,
   $\{0.62, 1.18, 1.85, 1.26, 1.31\}$~${\rm arcmin}^{-2}$ for five source bins,
   and values derived from our selection pipeline at KiDS-like depth,
   $\{0.21, 0.11, 0.06, 0.05, 0.05, 0.08\}$~${\rm arcmin}^{-2}$ for six lens bins.
   For the DES-like and LSST-like configurations, we uniformly rescale these per-bin values to match the
   reported or targeted total number densities of each survey{. We also rescale the tile size for these
   configurations according to the focal planes of the respective instruments,
   as detailed in Appendix~\ref{sec:impact_tile_size}.}

   The resulting totals are 6.22, 8.30, and 27.73~${\rm arcmin}^{-2}$ for sources, and
   0.56, 0.22, and 16.14~${\rm arcmin}^{-2}$ for lenses, in the KiDS-like, DES-like, and LSST-like setups, respectively.
   We note that these number densities enter the analysis only through the covariance matrix.
   The same galaxy bias values are used across these substantially different survey
   depths. This is unrealistic, since lens samples of different spatial density would have substantially
   different stellar masses and hence different bias \citep{Zehavi2011}. However, this
   does not affect our analysis greatly: we study the multiplicative, scale-dependent distortion
   of $w(\theta)$ induced by the $n(z)$ variation, which is largely decoupled from the
   overall bias amplitude.

   {The observing-condition maps and the detection emulator, hence, the galaxy
   selection and the level of spatial $n(z,\boldsymbol{\theta})$ variation that sources the
   effect, are held at the KiDS-like depth for all three configurations. We do not
   re-simulate the selection at the deeper DES- or LSST-like imaging, nor re-train the
   emulator for it. The three setups therefore differ only in the tile size and in the
   survey area and number density that enter the covariance.
   The absolute amplitude of the effect is determined by the KiDS-like variation level and our
   fiducial lens sample selection, with a simplified rescaling of the tile size for the DES-like and LSST-like cases. Notably, we do not attempt to match the levels of tile-to-tile inhomogeneity of observing conditions in these other surveys, i.e.~{we} somewhat counterfactually assume for the purpose of our tests that surveys of DES and LSST-like depth and sky area are performed with a similar observing strategy as the one of KiDS.}

   \subsection{PPD analysis on internal consistency}
   \label{ppd_analysis}

   In a cosmological analysis, the angular clustering of galaxies is typically
   combined with other two-point correlation functions, in particular the
   cross-correlation of ``lens'' galaxies with the shapes of background ``source''
   galaxies. Before such an analysis can be concluded, surveys commonly require a
   set of tests to be passed while still blinded to the cosmological results
   \citep{Asgari2021,DESY32022,DESY62026}. Perhaps the most important of these is a
   test for internal consistency: if the galaxy clustering and the other parts of the
   data vector cannot be described well by the same parameters within a single model, then it
   is not justifiable to combine them to constrain those parameters
   \citep[see the unblinding criteria of][]{DESY32022,DESY62026}. We test this with
   a posterior-predictive (PPD) internal-consistency test \citep{DouxPPD2021}: we
   condition on probes unaffected by the contamination, predict the clustering, and
   compare to the contaminated measurement. {We use galaxy--galaxy lensing to predict the clustering because lensing is
   less affected by the redshift variation \citep{HangLSST2024,Kong2026}; in our
   test, we add contamination only to $w(\theta)$.} We
   consider two conditioning sets: GGL alone, and GGL
   combined with cosmic shear, the latter pinning $\sigma_8$ and thereby breaking the
   $b_i$--$\sigma_8$ degeneracy that limits the GGL-only version. For each set we fit
   a conditioning-only chain with the same pipeline and covariance as in
   Sect.~\ref{sec:cosmo_setup}.

   $\chi^2$-type discrepancy statistics can bias the
   PPD $p$-value low even for perfectly consistent noisy data \citep{DouxPPD2021}.
   Therefore we characterize the
   test by the sampling distribution of its statistic under noise, using
   the simulation machinery that \citet{DouxPPD2021} introduce to calibrate their
   measured $p$-values. We draw $10^4$ noise realizations jointly
   across all probes, obtain each realization's posterior by {reweighting each sample of the conditioning-only chain by the ratio of
   the likelihood of the realization's conditioning data to that of the fiducial
   conditioning data, evaluated at the same parameter values}, and
   {compute its raw $p_j$. A second ensemble} adds the injected contamination
   $\Delta w$ (which leaves the conditioning probes unchanged) to the same noise
   realizations. The calibrated $\tilde p$ of a
   realization is the fraction of null realizations with smaller raw $p$
   \citep[Eq.~9 of][]{DouxPPD2021}.

   \begin{figure*}
      \centering
      \includegraphics[width=1\linewidth]{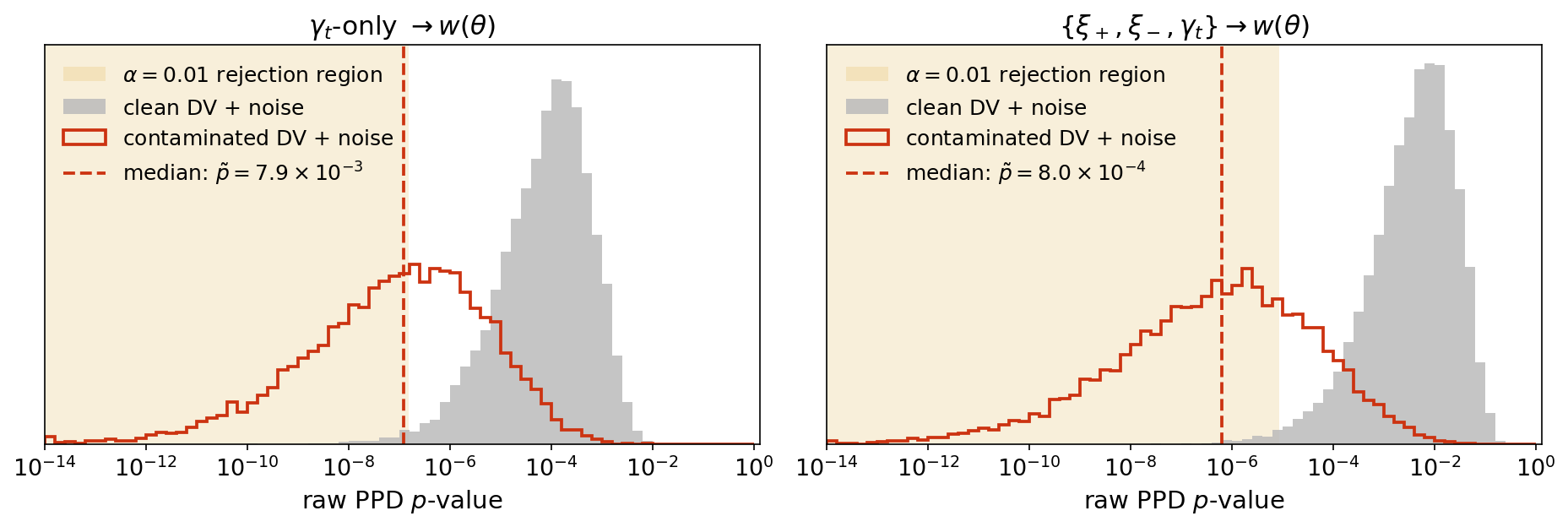}
      \caption{Sampling distributions of the raw PPD $p$-value for noisy data
      vectors, conditioning on GGL alone (\emph{left}) and on GGL plus cosmic shear
      (\emph{right}) {and predicting} $w(\theta)$. Gray: $10^4$ consistent realizations
      (clean fiducial vector plus noise across all probes); red: the same
      noise realizations with the $w(\theta)$ contamination {applied}. The raw statistic
      is strongly non-uniform even for consistent data (gray medians
      $\sim\!10^{-4}$ and $5\times10^{-3}$). The
      calibrated $\tilde p$ of a realization is {the integral over the gray histogram to its left}. The
      dashed line marks the median contaminated realization,
      $\tilde p=7.9\times10^{-3}$ (left) and $8.0\times10^{-4}$ (right). The shaded
      region (raw $p$ below the null $1\%$ quantile) is the $\alpha=0.01$ rejection
      region and contains $52\%$ (left) versus $73\%$ (right) of the contaminated realizations.}
      \label{fig:ppd_consistency}
   \end{figure*}

   The significance of the contamination depends strongly on the statistical
   power of the dataset considered. Here, we present results assuming an LSST-like
   covariance.
   Figure~\ref{fig:ppd_consistency} summarizes the result. Conditioning on GGL alone,
   $52\%$ of noise realizations deliver $p$-values lower than 1\% of null realizations
   (median calibrated $\tilde p = 7.9\times10^{-3}$). The test
   has useful but limited power, because GGL alone leaves the $b_i$--$\sigma_8$
   amplitude poorly constrained and the {predictive distribution} correspondingly broad. Adding cosmic shear to
   the conditioning set pins $\sigma_8$ and shrinks the predictive spread. The fraction
   rises to $73\%$ at $\alpha=0.01$
   ($89\%$ at $\alpha=0.05$), with a median $\tilde p = 8.0\times10^{-4}$, beyond
   the $\tilde p>0.01$ consistency criterion of \citet{DouxPPD2021}. {With GGL and cosmic shear as conditioning probes, the contaminated
   clustering fails the adopted internal-consistency criterion in $73\%$ of
   realizations. Those realizations would not pass this criterion for unblinding;
   a joint shift in cosmological and bias parameters does not fully absorb the contamination.}
   The corresponding DES-like result is shown in
   Appendix~\ref{sec:app_ppd_des}.

   \subsection{Shifts in cosmological parameters}
   \label{sec:cosmo_cosmo}

   {Having established in Sect.~\ref{ppd_analysis} that the contamination breaks
   the internal consistency between clustering and lensing for an LSST-like survey,
   we now quantify the parameter bias incurred when the effect is left unmodeled.}
   As shown in Sect.~\ref{sec:main_results}, the contamination inflates $w(\theta)$,
   most strongly at small angular scales and in high-redshift bins. A misspecified model
   can compensate {for this} by exploiting the {$b^2 \sigma_8^2$} degeneracy in the clustering signal.
   However, the galaxy--galaxy lensing signal $\gamma_t \propto b\,\sigma_8^2$, which
   is unaffected by the redshift distribution variation, anchors the combination
   {$b\,\sigma_8^2$}. The joint constraint therefore drives the model toward higher galaxy bias
   $b$ and {lower $\sigma_8$}. Because
   $h_0$ and $n_s$ are poorly constrained by this $2\times2$pt data combination,
   we hold them fixed at their fiducial values in the chains presented here. In
   Appendix~\ref{sec:app_relaxed_cosmology}, we free both parameters {for a more realistic setup}.

\begin{figure*}
      \centering
      \includegraphics[width=1\linewidth]{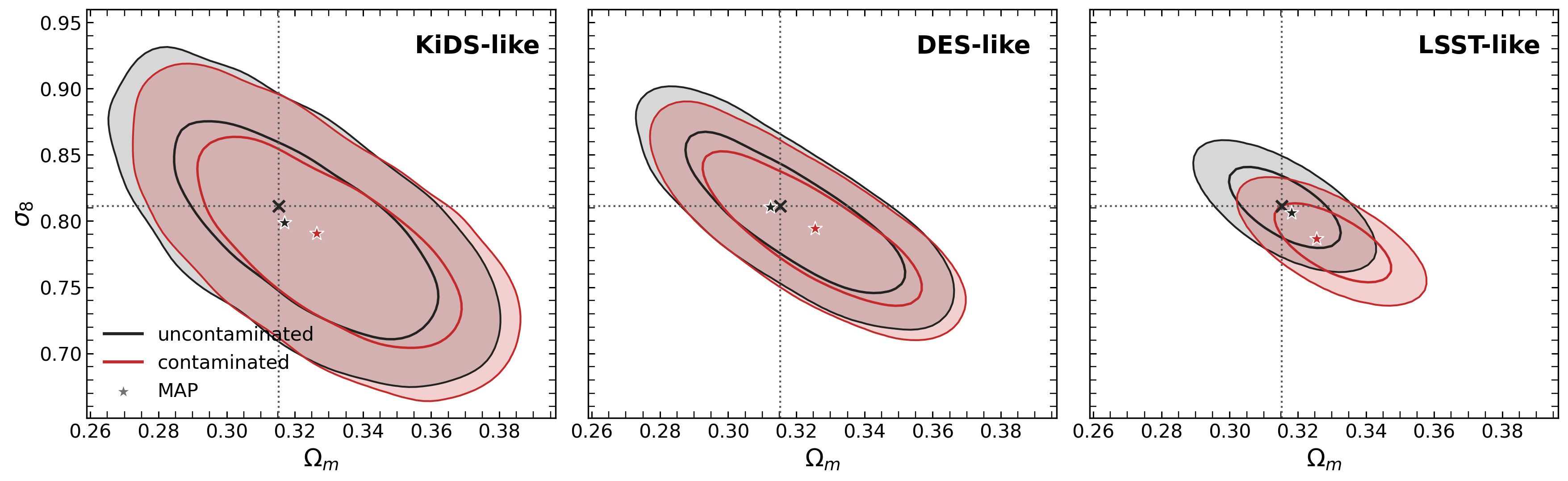}
      \caption{Joint posterior of $\Omega_m$ and $\sigma_8$ for the three survey
      configurations (KiDS-like, DES-like, LSST-like), comparing uncontaminated
      (black) and contaminated (red) data vectors. Dashed lines mark the fiducial
      parameter values. Contamination shifts the posterior along the
      $\Omega_m$--$\sigma_8$ degeneracy direction: $\sigma_8$ is biased low and
      $\Omega_m$ is biased high. The effect is sub-$1\sigma$ for KiDS-like and
      DES-like configurations, but reaches $\sim -1.7\sigma$ in $\sigma_8$
      for the LSST-like setup owing to its smaller statistical uncertainties and
      larger tile size. }
      \label{fig:omega_m_s8_fixed}
   \end{figure*}

   Fig.~\ref{fig:omega_m_s8_fixed} shows the joint posterior of $\Omega_m$ and
   $\sigma_8$ for the three survey configurations. For KiDS-like and DES-like setups
   the cosmological parameter shifts are well below $1\sigma$ ($\sigma_8$ shifts by
   $\sim -0.2\sigma$ and $\sim -0.3\sigma$, respectively) and remain within the
   statistical uncertainty. For the LSST-like setup, the shift in $\sigma_8$
   reaches $-1.7\sigma$ (from $0.8103$ to $0.7837$), and in $A_s$ reaches $-2.7\sigma$
   (from {$2.095\times10^{-9}$} to $1.903 \times 10^{-9}$); the smaller statistical uncertainties of
   a Stage-IV survey render the same absolute offset significantly more important.
   Notably, the shift in $\Omega_m$ is in the opposite direction ($+1.4\sigma$,
   from $0.3161$ to $0.3300$), since the contamination moves the posterior along the
   $\Omega_m$--$\sigma_8$ degeneracy axis. The derived parameter $S_8 \equiv \sigma_8
   (\Omega_m/0.3)^{0.5}$, {which is approximately constant along this degeneracy direction}, is
   correspondingly more stable: the LSST-like shift in $S_8$ is only $-1.2\sigma$
   (from $0.8314$ to $0.8217$), and sub-$0.3\sigma$ for KiDS-like and DES-like.

\begin{figure*}
      \centering
      \includegraphics[width=1\linewidth]{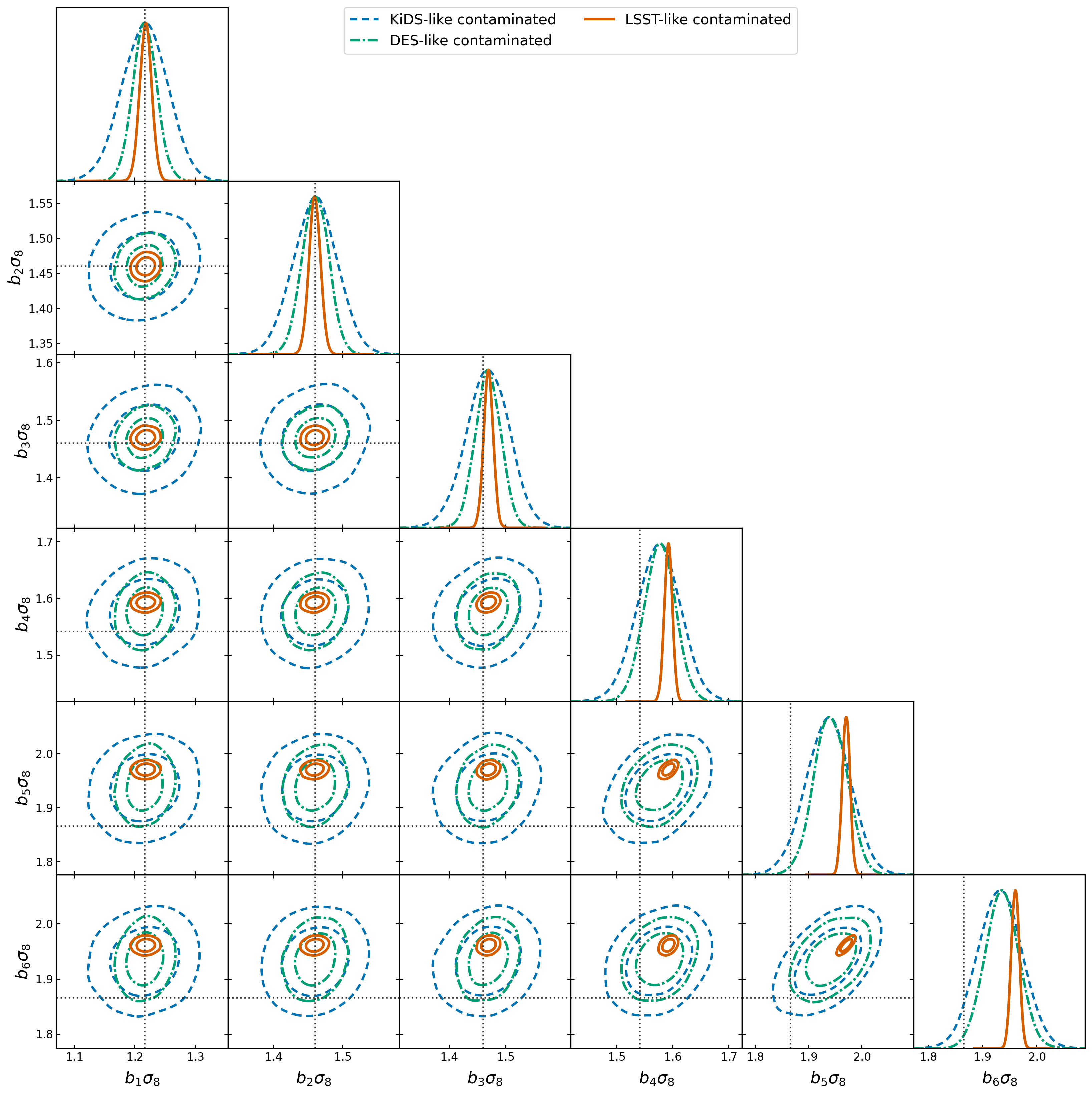}
      \caption{Triangle plot of $b_i\sigma_8$ for the six tomographic bins from the
      contaminated LSST-like chain (orange) overlaid on the uncontaminated chain
      (blue). Dashed lines mark the fiducial values. The bias parameters absorb the
      bulk of the excess clustering signal, with shifts increasing toward higher
      redshift bins; $b_5$ and $b_6$ are the most strongly affected. Cosmological
      parameters $\Omega_m$ and $\sigma_8$ (shown in Fig.~\ref{fig:omega_m_s8_fixed})
      are less strongly shifted by comparison.}
      \label{fig:bsig8_fixed}
   \end{figure*}

   The bulk of the excess clustering signal is absorbed by the galaxy bias
   parameters, as shown in Fig.~\ref{fig:bsig8_fixed}. {This preferential absorption
   by bias rather than cosmological parameters occurs because the contamination is
   concentrated in the high-redshift bins (Sect.~\ref{sec:main_results}).} The corresponding bias parameters
   $b_5$ and $b_6$ can absorb this excess without disturbing the lower-redshift bins. As a result, the bias shifts increase strongly with redshift: in the
   LSST-like case, $b_1$ shifts by $\sim +1.6\sigma$ while $b_5$ and $b_6$ shift by
   $\sim +4.3\sigma$ and $\sim +4.1\sigma$, respectively. For KiDS-like and DES-like
   the bias shifts reach $\sim +0.8\sigma$ and $\sim +1.1\sigma$ for $b_5$.

   We note that the effect on cosmological parameters would likely increase if either a continuity of bias across redshift bins was imposed or additional freedom was given to the redshift evolution of the structure amplitude.

   In Table~\ref{tab:fixed-marginal-constraints} we present the full marginal
   posterior constraints, together with goodness-of-fit values evaluated at the
   maximum-posterior point of each chain. The joint $\chi^2$ is computed with the
   full covariance of the combined $w(\theta)$ and $\gamma_t$ data vector.
   We quote standalone values for the clustering and galaxy--galaxy lensing
    sub-vectors, using the corresponding sub-blocks of the covariance.
   Because the data vectors are noiseless, the raw $\chi^2$ values should not
   be interpreted as draws from a $\chi^2$ distribution.
   For {the KiDS-like configuration}, the contaminated and uncontaminated fits remain almost unaffected.
   The DES-like configuration shows mild degradation, particularly when clustering is involved.
   For the LSST-like data, however, contamination increases the joint statistic
   by $\Delta\chi^2_{\rm joint}=25.26$ for our 449-entry data vector, with
   increases of $19.88$ and $5.87$ for the 69-entry $w(\theta)$ and 380-entry GGL
   sub-vectors, respectively.
   The clustering term reflects the residual scale- and redshift-dependent
   mismatch left after the model adjusts the bias parameters, while the GGL term
   reflects the inflated biases required to fit the excess clustering, consistent
   with the clustering--lensing inconsistency identified by the PPD test
   (Sect.~\ref{ppd_analysis}). {We stress that the LSST-like setup shares the same
   KiDS-like selection depth as the others (Sect.~\ref{sec:cosmo_setup}). Its far poorer
   fit reflects its tighter covariance, which makes the similar absolute contamination more significant.}

\begin{table*}
      \centering
      \small
      \setlength{\tabcolsep}{4pt}
      \caption{\protect{Marginal posterior constraints of chains fixing $n_s$ and $h_0$ to
      their fiducial values. Parameter entries give the posterior mean and marginal
      $68\%$ standard deviation. The $\chi^2$ rows are evaluated at the
      maximum-posterior sample {and for noiseless data. ``Joint''} uses the full $w(\theta)+\gamma_t$
      covariance, while the $w(\theta)$ and GGL rows use the corresponding
      covariance sub-blocks {only}.}}
      \label{tab:fixed-marginal-constraints}
      \begin{tabular}{l c cc cc cc}
      \toprule
      Parameter & Truth & \multicolumn{2}{c}{KiDS-like} & \multicolumn{2}{c}{DES-like} & \multicolumn{2}{c}{LSST-like} \\
      \cmidrule(lr){3-4}\cmidrule(lr){5-6}\cmidrule(lr){7-8}
      &  & Uncontam. & Contam. & Uncontam. & Contam. & Uncontam. & Contam. \\
      \midrule
      $\Omega_m$ & $0.3153$ & $0.3227\pm0.0238$ & $0.3291\pm0.0238$ & $0.3194\pm0.0193$ & $0.3238\pm0.0191$ & $0.3161\pm0.0097$ & $0.3300\pm0.0101$ \\
      $\sigma_8$ & $0.8113$ & $0.7946\pm0.0508$ & $0.7838\pm0.0495$ & $0.8046\pm0.0358$ & $0.7943\pm0.0346$ & $0.8103\pm0.0168$ & $0.7837\pm0.0161$ \\
      $S_8$ & $0.8317$ & $0.8221\pm0.0378$ & $0.8191\pm0.0371$ & $0.8290\pm0.0199$ & $0.8239\pm0.0193$ & $0.8314\pm0.0081$ & $0.8217\pm0.0079$ \\
      $\Omega_b$ & $0.0493$ & $0.0517\pm0.0107$ & $0.0519\pm0.0108$ & $0.0510\pm0.0099$ & $0.0515\pm0.0098$ & $0.0496\pm0.0051$ & $0.0526\pm0.0053$ \\
      {$10^9 A_s$} & $2.10$ & $2.018\pm0.300$ & $1.909\pm0.282$ & $2.069\pm0.178$ & $1.987\pm0.167$ & $2.095\pm0.080$ & $1.903\pm0.071$ \\
      $b_1$ & $1.50$ & $1.533\pm0.102$ & $1.559\pm0.102$ & $1.514\pm0.069$ & $1.535\pm0.069$ & $1.502\pm0.033$ & $1.556\pm0.034$ \\
      $b_2$ & $1.80$ & $1.843\pm0.119$ & $1.870\pm0.118$ & $1.818\pm0.081$ & $1.842\pm0.080$ & $1.803\pm0.037$ & $1.863\pm0.038$ \\
      $b_3$ & $1.80$ & $1.844\pm0.126$ & $1.881\pm0.126$ & $1.818\pm0.084$ & $1.853\pm0.084$ & $1.803\pm0.037$ & $1.876\pm0.037$ \\
      $b_4$ & $1.90$ & $1.946\pm0.133$ & $2.017\pm0.134$ & $1.919\pm0.091$ & $1.989\pm0.091$ & $1.903\pm0.038$ & $2.033\pm0.040$ \\
      $b_5$ & $2.30$ & $2.356\pm0.158$ & $2.480\pm0.162$ & $2.322\pm0.108$ & $2.449\pm0.111$ & $2.303\pm0.046$ & $2.515\pm0.049$ \\
      $b_6$ & $2.30$ & $2.357\pm0.159$ & $2.475\pm0.163$ & $2.322\pm0.108$ & $2.442\pm0.110$ & $2.303\pm0.046$ & $2.502\pm0.049$ \\
      \midrule
      {$\chi^2_{\rm joint}$} & {--} & {$0.732$} & {$1.040$} & {$0.203$} & {$1.336$} & {$0.690$} & {$25.953$} \\
      {$\chi^2_{\rm GGL}$} & {--} & {$0.182$} & {$0.072$} & {$0.101$} & {$0.324$} & {$0.117$} & {$5.990$} \\
      {$\chi^2_{w(\theta)}$} & {--} & {$0.662$} & {$0.930$} & {$0.098$} & {$1.071$} & {$0.569$} & {$20.444$} \\
      \bottomrule
      \end{tabular}
      \end{table*}

   \subsection{Dependence on survey configuration}
   \label{sec:cosmo_survey}

   {The results above illustrate a qualitative difference between the three survey
   configurations.} For KiDS-like and DES-like setups,
   the contamination in cosmological parameters is sub-$0.5\sigma$ and comparable
   to the statistical floor. {For the LSST-like setup, the shifts in $\sigma_8$
   ($1.7\sigma$) and $A_s$ ($2.7\sigma$) exceed the statistical error, so the
   effect must be accounted for to keep biases below that error.}

   {This qualitative change has two contributing origins.} The first is simply the
   smaller statistical uncertainties of a Stage-IV survey: the same absolute parameter
   shift {is more significant} when the posterior is narrower. The second
    origin is the larger tile size of LSST ({$\sim 9\,\mathrm{deg}^2$})
   compared to KiDS ($\sim 1\,\mathrm{deg}^2$). As shown in
   Appendix~\ref{sec:impact_tile_size}, larger tiles imprint the redshift-distribution
   variation onto larger angular scales.

   While we have calibrated our three survey configurations to real survey
   specifications, several gaps remain between our model and an actual analysis.
   {Of the survey properties we vary, tile size directly sets the angular scale of
   the contamination, while number density and sky coverage enter only through the
   covariance. Beyond these, deeper surveys can also
   shift the magnitude distribution of selected galaxies,
   changing the sensitivity of the selection even at fixed
   observing-condition variation amplitude. We hold the amplitude of observing-condition variation at the KiDS-like
   level while increasing the tile size, which changes its spatial coherence scale.
   In practice, the multi-visit strategy of LSST
   partially averages down systematic gradients within a pointing (as noted in
   Sect.~\ref{systematics}), and the actual variation amplitude will depend strongly
   on the adopted survey strategy. A faithful, survey-specific
   quantification would therefore require the actual systematic maps, selection
   pipeline, and covariance model of each survey.} Finally, because the
   contamination shifts the growth amplitude, we extend the inference to a
   dynamical dark-energy ($w_0w_a$CDM) model that frees the low-redshift growth
   history, testing whether the bias is partly absorbed into $w_0$/$w_a$ rather
   than $\sigma_8$. As shown in Appendix~\ref{sec:app_w0wa}, for an LSST-like
   survey it largely is: the $\sigma_8$ bias of the $\Lambda$CDM chains is
   rerouted into a spurious $\sim1\sigma$ preference for $w_a>0$.

   \section{Conclusions}
   \label{sec:conclusions}

   Standard photometric clustering analyses assume an isotropic redshift
   distribution $n(z)$, while in practice this assumption is broken by spatially
   varying observational systematics. The theoretical framework and qualitative
   implications of this anisotropy have been laid out in the literature
   \citep{LizancosSpatial2023, Kong2026}, but a quantitative end-to-end
   assessment at a realistic survey level has been missing. {We quantify this impact by forward-modeling how systematics maps affect
   detection, photometry, tomographic binning, and the clustering data vector.
   We then infer cosmological parameters while ignoring the effect and measure
   the resulting bias.}

   In our tomographic setup, modeling $w(\theta)$ with a constant $\bar{n}(z)$
   under-predicts the two-point correlation by $\sim$10\% at small scales
   ($\theta \lesssim 10$\,arcmin) in the {$z\in(0.85,0.95]$ and
   $z\in(0.95,1.05]$ bins, by $\sim$7\% in $z\in(0.70,0.85]$}, by $\sim$2\% in
   $z\in(0.55,0.70]$, and negligibly at lower redshifts.  The characteristic
   scale of the contamination is set by survey geometry and observing
   strategy, which we jointly summarize as the tile size: larger tiles
   imprint the boosted $w(\theta)$ on correspondingly larger angular
   separations.

   {We examine the consequences for cosmological inference.} Before
   interpreting parameter shifts from a joint fit, we first ask whether the
   contaminated clustering signal remains internally consistent with lensing. A
   posterior-predictive test shows a probe-to-probe inconsistency under
   LSST-like noise. In $52\%$ of
   realizations, contaminated clustering appears {inconsistent at the $\tilde{p}<1\%$ level} given GGL (median calibrated
   $\tilde p\simeq8\times10^{-3}$).
   Extending the conditioning set to cosmic shear breaks {the $b_i$--$\sigma_8$ degeneracy} and {causes
   $73\%$ of realizations to fail the $\tilde p>0.01$ consistency criterion.}

   {Having established this inconsistency, we quantify the parameter bias incurred
   when the effect is nevertheless left unmodeled.} The contamination is partially
   degenerate with the matter-fluctuation {amplitude $\sigma_8$}, but its
   strong redshift dependence allows it to be preferentially absorbed by the
   redshift-dependent linear galaxy-bias parameters of the highest-redshift bins.
   We considered three survey configurations spanning the relevant range of
   statistical power and tile size: KiDS-like, DES-like, and LSST-like. For the
   KiDS-like and DES-like cases, the shifts in $\Omega_m$, $\sigma_8$, and $S_8$
   all remain below $0.5\sigma$, and the goodness-of-fit is essentially unchanged.
   For the LSST-like case, where statistical uncertainties are smaller and the tile
   is larger, the shifts in $\sigma_8$ and $A_s$ reach $1.7\sigma$ and $2.7\sigma$,
   respectively, {while} $h_0$ and $n_s$ are held fixed.

   Several simplifications limit the direct applicability of our
   quantitative results. Most importantly, we did not use actual survey
   systematics maps but only parametric models broadly calibrated to KiDS
   data. {These models estimate the amplitude of the effect for KiDS-like observing
   conditions. A survey-specific calibration would require the systematics maps,
   selection pipeline, and covariance model of the target survey.}

   {In our KiDS-like and DES-like configurations, spatial variation in $n(z)$
   is a subdominant but non-negligible systematic. In our LSST-like configuration,
   which retains KiDS-like selection and observing-condition variation, it produces
   significant parameter biases and frequent failures of the internal-consistency
   test. For this configuration, scale cuts and density-based random-catalog
   corrections alone are insufficient. Modeling or correcting the remaining effect
   is needed to meet the adopted consistency criterion and avoid parameter bias.} {Given {image simulations} that can be emulated as in this work, the forward model developed here
   can predict the contamination needed to correct the data.}

   \begin{acknowledgements}
      This work was supported by the Excellence Cluster ORIGINS,
      funded by the Deutsche Forschungsgemeinschaft (DFG, German Research Foundation)
      under Germany’s Excellence Strategy – EXC-2094/2 – 390783311.
      ZZ acknowledges support from the German Academic Exchange Service (DAAD).
      YW acknowledges the support of the China Scholarship Council program.
   \end{acknowledgements}

   \bibliographystyle{bibtex/aa}
   \bibliography{refs}

   \begin{appendix}

   \section{Impact of tile size}
   \label{sec:impact_tile_size}

   {A ``tile'' denotes a contiguous sky subregion whose imaging shares a common
   observing epoch and instrumental setup. Many of the dominant observational
   systematics vary in time: atmospheric seeing, sky brightness, and transparency
   change from night to night. In a single-epoch
   survey the exposures making up a tile are acquired over a short time span, so these
   conditions are nearly constant within the tile and the resulting systematics are
   spatially coherent on the tile scale. Larger tiles therefore imprint systematic
   variation on correspondingly larger angular scales. In multi-visit surveys this
   coherence is partly still present, but washed out by partially independent observing conditions between subsequent visits of the same field.}

\begin{figure*}
      \centering
      \includegraphics[width=1\linewidth]{
         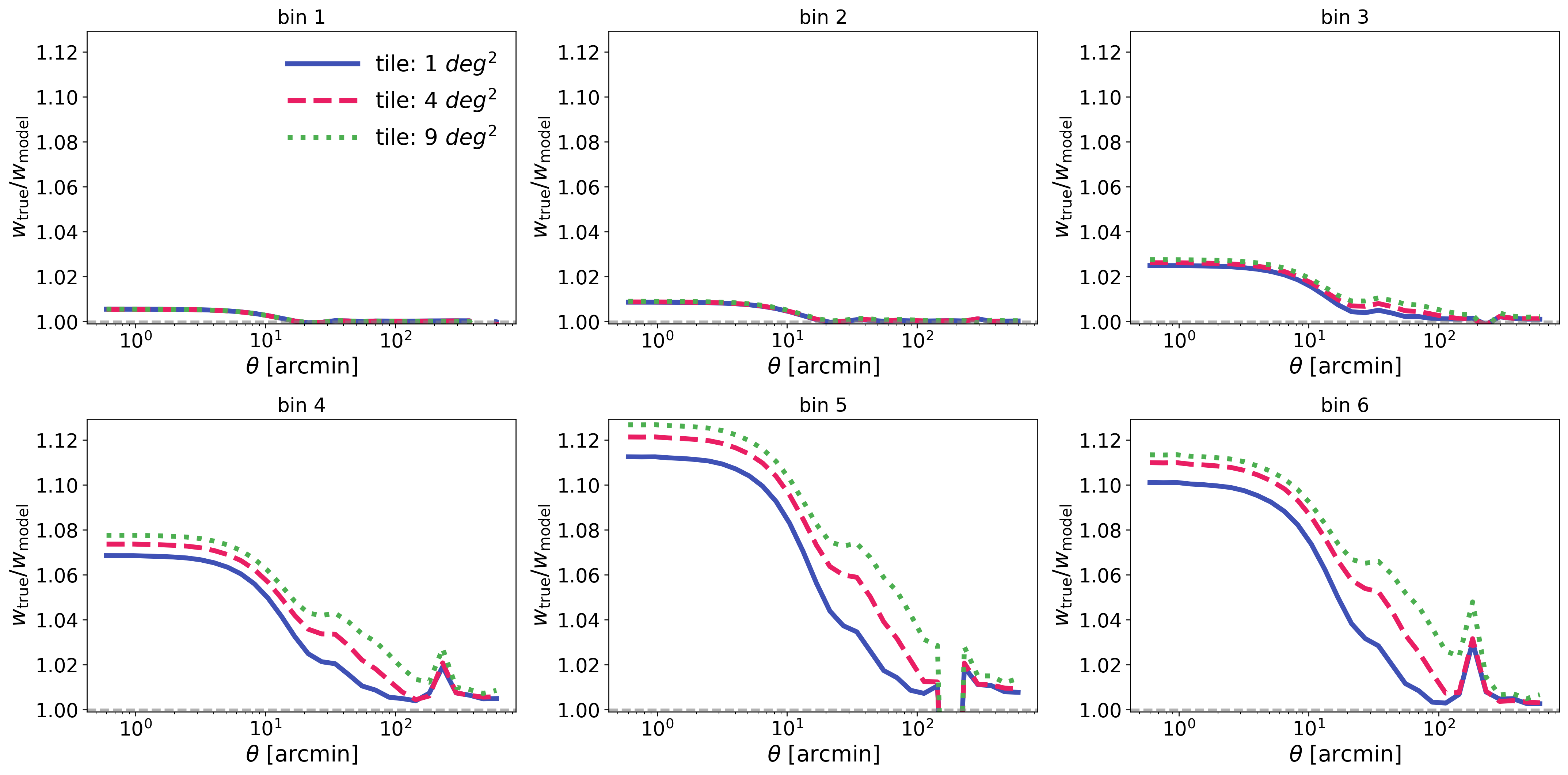
      }
      \caption{Ratio $w_{\rm true}/w_{\rm model}$ as a function of angular
      separation $\theta$ for three tile sizes: 1\,deg$^2$ (KiDS-like, solid),
      {4\,deg$^2$ (DES-like, dashed), and 9\,deg$^2$ (LSST-like, dotted)}. Each panel
      corresponds to one tomographic bin. Larger tiles imprint the bias on larger
      angular scales and amplify its overall amplitude.}
      \label{fig:ratio_tile}
   \end{figure*}

   {Fig.~\ref{fig:ratio_tile} shows the effect for three representative tile sizes
   of 1, 4, and 9\,deg$^2$ (1, 2, and 3\,deg on a side), spanning the range from the
   single-epoch KiDS regime toward the larger fields of view of DES \citep{DESCamera2015,DESDataSet2018} and LSST \citep{IvezicLSST2019}. This is
   a controlled scan of the coherence scale at fixed, KiDS-like variation amplitude:
   we hold the amplitude of the systematic variation constant and stretch only the
   tile size, keeping the tiles square, whereas real surveys use a variety of
   footprint geometries (DES, for example, observes in hexagon-like subareas).
   We return to this in Sect.~\ref{sec:cosmo_survey}.}

   As shown, larger tile sizes amplify the effect. For bin 6 at the smallest separation
   ($\approx 1$\,arcmin), the bias increases from 10\% for 1\,deg$^2$ to 15\% for
   {9\,deg$^2$}, with similar trends in other bins. The effect is even more pronounced
   at larger separations: for bin 6, a tile size of {9\,deg$^2$} raises the bias from
   nearly zero to $\sim$4\% at 100\,arcmin.

   \section{Impact of sample selection}
   \label{sec:impact_sample_selection}

   The spatial anisotropy in $w(\theta)$ has two distinct sources. For a sample
   selected only with a {loose} SNR threshold, spatial variation in the redshift
   distribution arises primarily from detection: worse
   observing conditions preferentially remove faint
   galaxies, retaining a sample skewed toward lower redshifts. This effect is most
   pronounced when considering the full redshift range. In a tomographic analysis,
   where galaxies are confined to narrow redshift intervals, the net re-weighting
   between the two ends of the distribution is weaker, and the second source,
   perturbation of bin assignment by observational systematics, becomes more
   important.

   {{The two selections differ in what they pin down.} A {loose} SNR cut
   acts on the observed, condition-rescaled photometry itself: in any pixel,
   galaxies are detected down to the limit set by the local conditions, so the
   surviving sample always sits at a similar rescaled photometric quality
   regardless of those conditions.
   They therefore share a similar distribution of $\tilde{f}$, and hence of the photo-$z$ scatter $\sigma_z$.
   Because tomographic bin assignment depends on
   this rescaled photometry, the binning response is weakly condition-dependent and
   the photo-$z$-induced $n(z,\boldsymbol{\theta})$ variation is minor. The detected
   galaxy population still varies across the sky because poorly observed pixels lose
   their faintest, highest-redshift galaxies, but this detection effect acts
   over the full redshift range rather than within a bin. The MagLim selection
   behaves oppositely. Although it is applied to the observed magnitude $\hat{m}_r$,
   the lens sample lies well above the detection limit, so its cut is,
   to good approximation, a selection on intrinsic magnitude. The intrinsic
   composition of the overall selected sample is then nearly the same across pixels.
   What varies with conditions is instead the quality of the observed photometry of
   those galaxies, affecting their photo-$z$ scatter $\sigma_z$ and hence the bin assignment.
   The photo-$z$-induced variation is correspondingly larger.}%

\begin{figure*}
      \centering
      \includegraphics[width=1\linewidth]{
         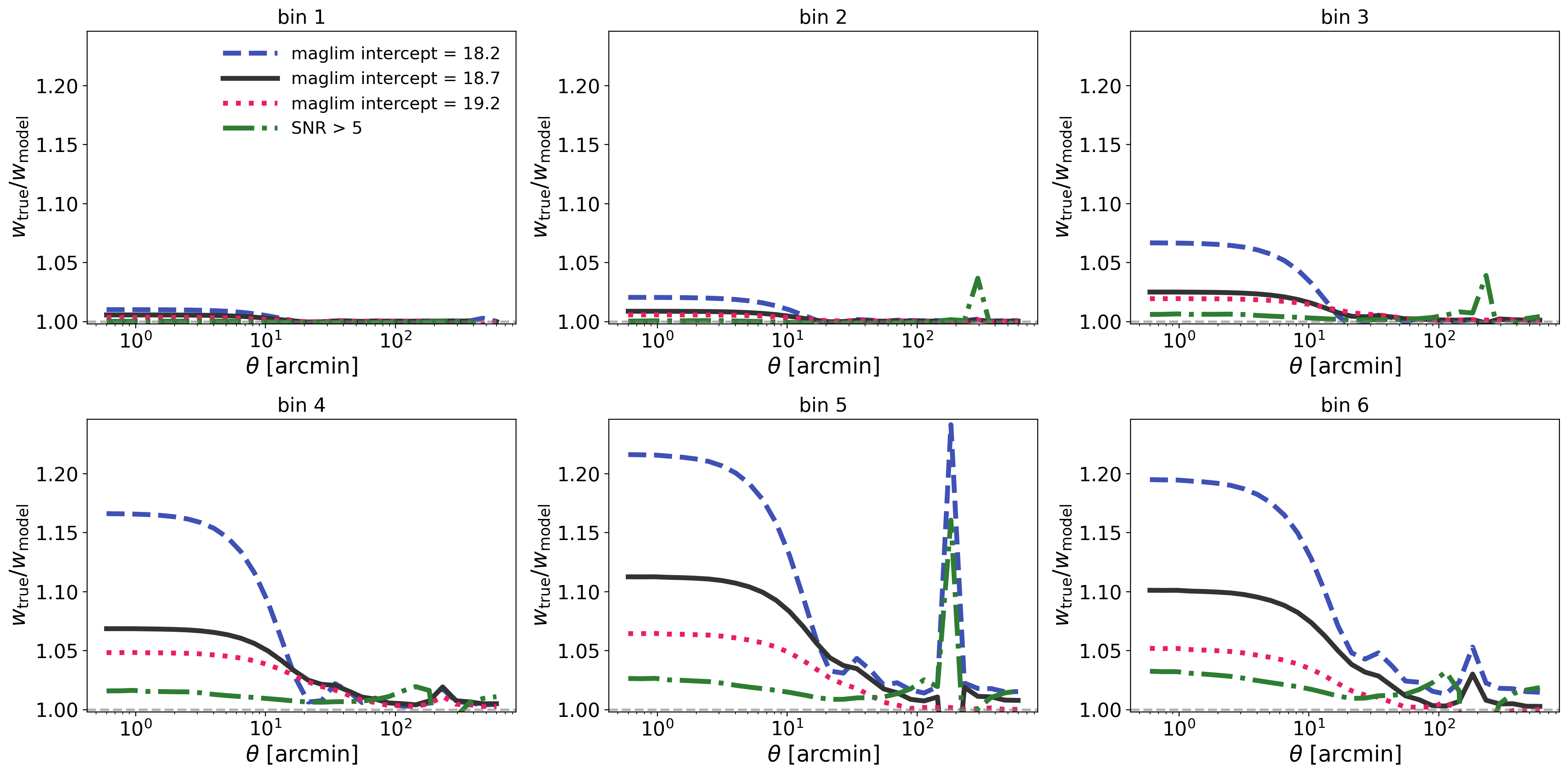
      }
      \caption{Ratio $w_{\rm true}/w_{\rm model}$ for different sample selection
      criteria. Lines show three MagLim-like cuts with varying intercept
      ({$\mathrm{maglim}_1 = 18.2$}, $18.7$, $19.2$; tight to loose) and a {loose} SNR\,$>$\,5
      cut. Each panel corresponds to one tomographic bin.
      Tighter magnitude cuts produce stronger biases, particularly in high-redshift
      bins, as the selected sample's photometric quality becomes more sensitive to
      local observing conditions.}
      \label{fig:ratio_selection}
   \end{figure*}

   Fig.~\ref{fig:ratio_selection} compares the $w(\theta)$ ratio for several
   selection choices: different magnitude thresholds for our MagLim-like cut and a
   {loose} SNR cut. Tighter cuts
   produce stronger variations. For the MagLim selection, the $w(\theta)$ impact for the
   highest-redshift bin drops from $\sim$20\% for {$\mathrm{maglim}_{1}=18.2$} to $\sim$5\% for
   {$\mathrm{maglim}_{1}=19.2$}, while the SNR\,$>$\,5 cut yields only $\sim$3\%. This is
   consistent with the arguments above: {looser cuts retain more galaxies near the
   detection limit, whose rescaled photometry is pinned close to the threshold, so the
   surviving sample shows a similar photometric-uncertainty distribution regardless of
   where on the sky it was observed}.

   \section{The impact of defining $\Bar{n}(z)$}
   \label{sec:impact_n_def}

   As shown in Eq.~\ref{term2}, {the shift term} is non-zero only when
   \begin{equation}
      \langle \epsilon(z,\boldsymbol{\theta}) \rangle \neq 0.
   \end{equation}
   It arises because {the detected }galaxy number density varies spatially,
   as expected given the respective observing conditions, and sky subregions
   with higher density contribute proportionally more to the global redshift
   distribution. {Standard analyses routinely account for density variation but often give
   little attention to how $\bar{n}(z)$ is defined when the density varies.} Specifically, $\bar{n}(z)$ can be estimated from the raw galaxy sample
   or from galaxies weighted inversely by the local number density expected based on
   the respective observing conditions, that is, an area-averaged estimate {\citep[e.g.,][]{DESlens2022}}.
   {The shift term} vanishes in the latter case.

\begin{figure}
      \centering
      \includegraphics[width=1\linewidth]{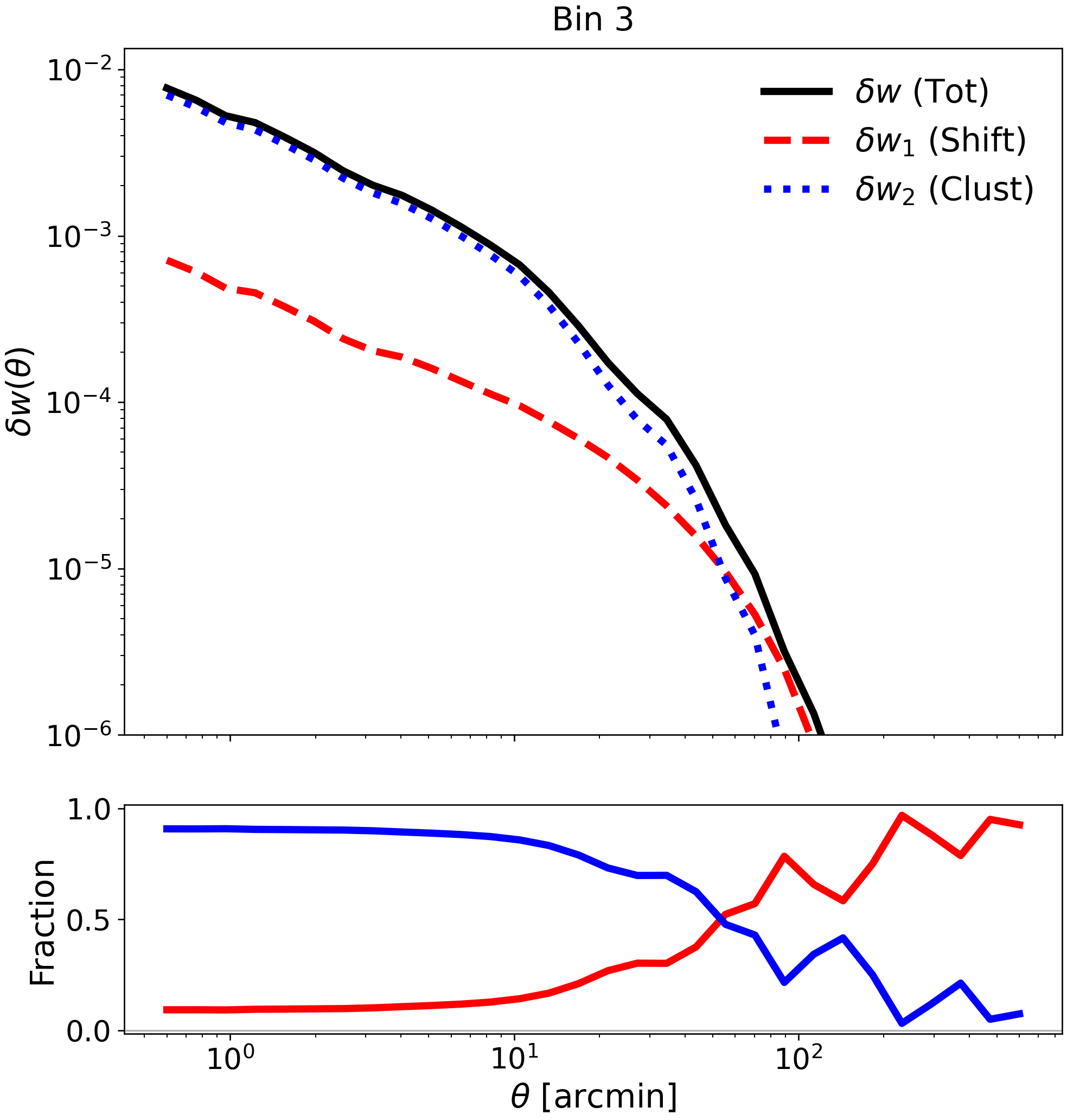}
      \includegraphics[width=1\linewidth]{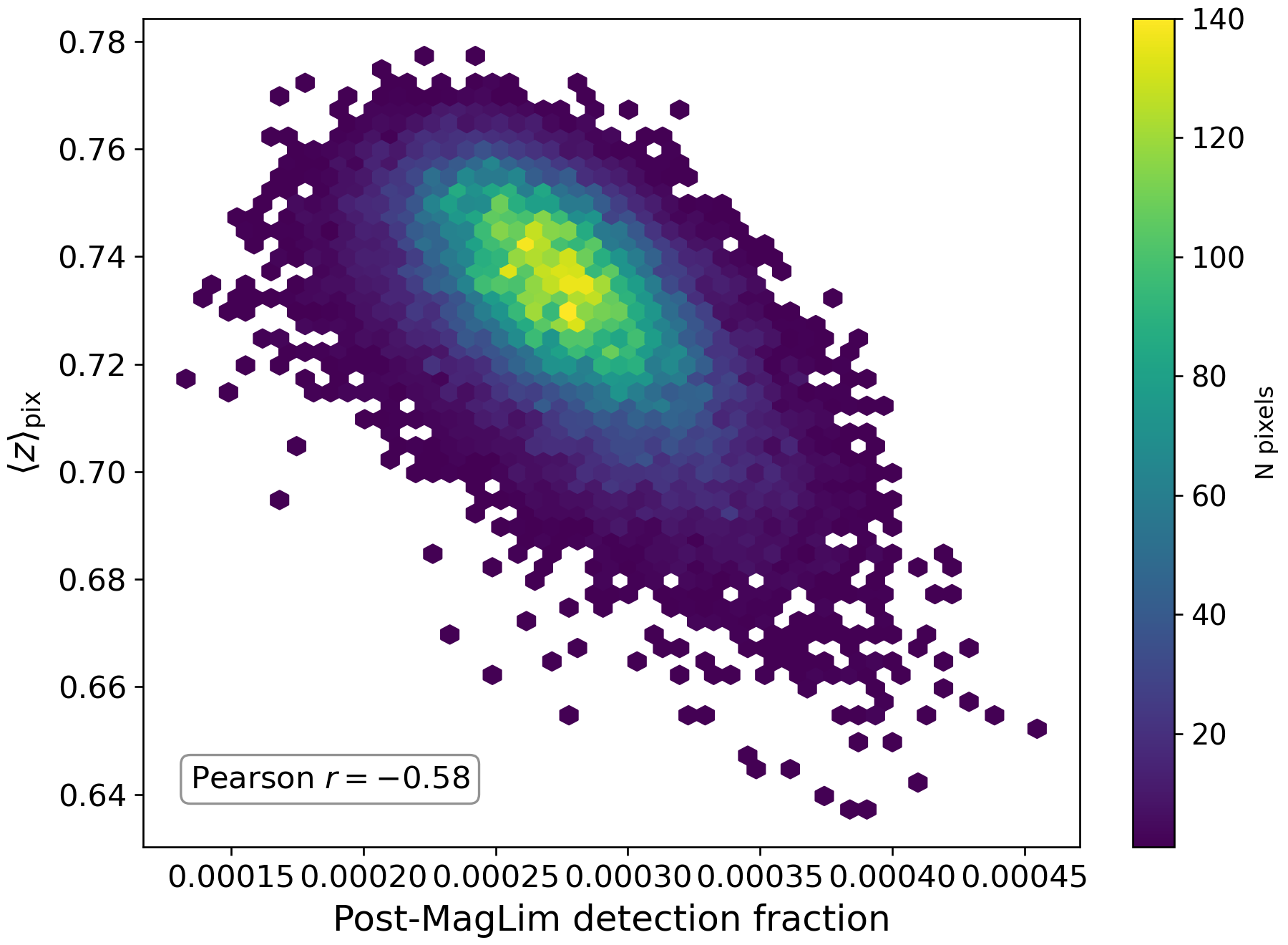}
      \caption{\textit{Top:} Contributions to the total bias $\Delta w(\theta) =
      w_{\rm true} - w_{\rm model}$ for tomographic Bin 3, decomposed into {the shift term}
      (``Shift'', arising from $\langle\epsilon\rangle \neq 0$) and {the clustering term} (``Clust'',
      arising from spatial correlation of $n(z,\boldsymbol{\theta})$), along with
      their sum (``Tot''). {The clustering term} dominates at small separations, while {the shift term} takes
      over at $\theta \gtrsim 50$\,arcmin. \textit{Bottom:} Pixel-averaged true mean
      redshift within Bin 3 as a function of the post-MagLim detection fraction per
      pixel. Pixels with higher detection fraction have systematically lower mean
      redshift (Pearson $r = -0.58$), driving the asymmetric shape of
      $\langle\epsilon(z)\rangle$ that gives {the shift term} its positive sign.}
      \label{fig:delta_w_terms}
   \end{figure}

   {Using the count-weighted $\bar{n}(z)$ estimate, in}
   Fig.~\ref{fig:delta_w_terms} we compare {the shift term and the clustering term} (labeled ``Shift''
   and ``Clust'' in the plot respectively), as well as the total difference in
   $w(\theta)$ (``Tot''). {For Bin 3 (other bins behave similarly), the shift term stays about an order of
   magnitude below the clustering term out to $\theta\simeq50$\,arcmin, beyond which
   it dominates.

   The shift term is not positive by construction. Its sign is set by the integrand
   $\bar{n}^{2}(z)\,\langle\epsilon(z)\rangle$, which is negative on the low-redshift
   side of the bin and positive on the high-redshift side. The positive lobe is the
   larger of the two in our case, hence the net term is positive.

   The lower panel of Fig.~\ref{fig:delta_w_terms} provides an understanding of this trend. Within Bin 3, better-observed
   pixels (higher post-MagLim detection fraction) have a \emph{lower} mean true redshift
   (Pearson $r=-0.58$). The reason is photo-$z$ scatter, which is larger in poorly observed
   pixels. Since the underlying $\mathrm{d}N/\mathrm{d}z$ rises across Bin 3, this scatter
   carries more galaxies into the bin from above than from below, so poorly observed pixels
   pick up high-redshift interlopers that raise their mean redshift. Well-observed pixels
   keep accurate redshifts near the nominal range, and hence a lower mean.

   Well-observed pixels also contain more galaxies, so they dominate the count-weighted
   $\bar{n}(z)$ and pull it toward low redshift relative to the area-averaged
   $\langle n(z)\rangle$. The difference $\langle n(z)\rangle-\bar{n}(z)$ is therefore
   positive on the high-redshift side, where $\bar{n}(z)$ peaks, giving a net positive
   shift term.}

      \section{Relaxed cosmology}
      \label{sec:app_relaxed_cosmology}

      In the fiducial inference test of Sect.~\ref{sec:cosmo_setup}, $h_0$
      and $n_s$ are fixed because photometric galaxy clustering and galaxy--galaxy
      lensing constrain them poorly. We now repeat the same test with both parameters
      free, while leaving the data vectors, covariance, likelihood, and scale cuts
      unchanged. {Freeing them mainly adds degeneracy volume rather than a direct
      response to the amplitude-like contamination.}

      The marginalized constraints are listed in
      Table~\ref{tab:relaxed-marginal-constraints}. Throughout this appendix,
      shifts are quoted as the difference between the contaminated and
      uncontaminated posterior means divided by the contaminated-chain marginal
      $68\%$ width. The direction of the shifts is the same as in the fixed
      chains: contaminated data prefer larger $\Omega_m$, lower $\sigma_8$ and
      lower $S_8$. The formal significance is reduced, however, because the
      relaxed chains are broader and because some absolute shifts are smaller.
      For the KiDS-like and DES-like configurations, the shifts in $\sigma_8$ are
      only $-0.10\sigma$ and $-0.25\sigma$, respectively, while the corresponding
      shifts in $S_8$ are $-0.05\sigma$ and $-0.23\sigma$.

\begin{figure*}
            \centering
            \includegraphics[width=1\linewidth]{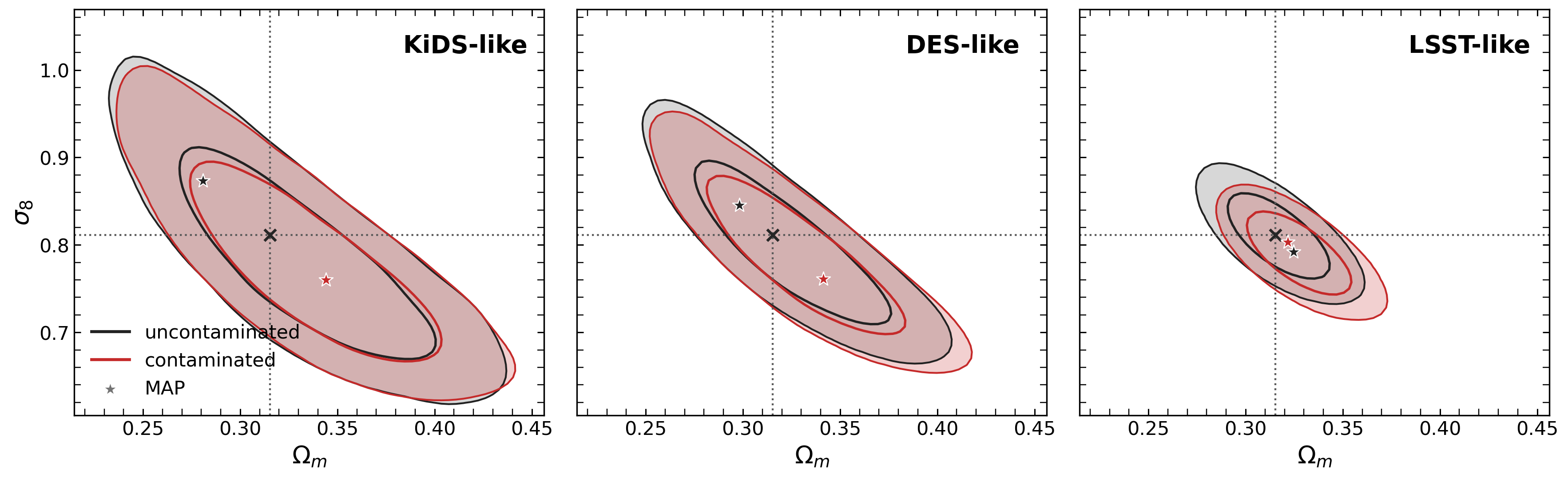}
            \caption{\protect{Joint posterior of $\Omega_m$ and {$\sigma_8$} for the relaxed
            chains, in which $h_0$ and $n_s$ are free. Compared with the fixed
            chains, the posterior contours are broader and the contamination-induced
            shifts are smaller in units of the marginal uncertainty. The LSST-like
            configuration remains the most affected, while the KiDS-like and
            DES-like configurations remain consistent within their statistical
            uncertainties.}}
            \label{fig:omega_m_s8_relaxed}
         \end{figure*}

      The LSST-like configuration remains the most affected case, as shown in
      Fig.~\ref{fig:omega_m_s8_relaxed}. In the relaxed chains, $\Omega_m$ shifts
      from $0.3167$ to $0.3275$ ($+0.70\sigma$), $\sigma_8$ shifts from $0.8100$
      to $0.7902$ ($-0.75\sigma$), and $S_8$ shifts from $0.8314$ to $0.8247$
      ($-0.60\sigma$). These shifts are smaller than in the fixed-parameter
      chains, where the corresponding significances are approximately
      $+1.4\sigma$, $-1.7\sigma$, and $-1.2\sigma$. Neither $h_0$ nor $n_s$
      shows a significant contamination-induced displacement:
      in the LSST-like case, $h_0$ shifts by $-0.33\sigma$ and $n_s$ by
      $+0.46\sigma$, with still smaller shifts for KiDS-like and DES-like
      configurations.

      {Freeing $h_0$ and $n_s$ therefore reduces the apparent cosmological
      bias in units of the marginalized posterior width, but it does not remove
      the underlying mismatch.} The excess clustering is still absorbed primarily
      by the galaxy-bias parameters, especially in the high-redshift bins. In the
      relaxed LSST-like chain, $b_5$ increases from $2.306$ to $2.505$
      ($+2.9\sigma$), and $b_6$ increases from $2.306$ to $2.492$
      ($+2.8\sigma$). The goodness-of-fit values in
      Table~\ref{tab:relaxed-marginal-constraints} show the same pattern as the
      fixed chains. The relaxed LSST-like joint $\chi^2$ increases from $1.05$
      to $25.10$ when the contamination is included, with standalone contributions
      increasing from $0.73$ to $18.00$ for $w(\theta)$ and from $0.42$ to
      $6.96$ for GGL. By contrast, the KiDS-like and DES-like configurations
      remain close to $\chi^2\simeq 1$ after contamination.

      \begin{table*}
      \centering
      \small
      \setlength{\tabcolsep}{4pt}
      \caption{\protect{Marginal posterior constraints for the relaxed chains,
      in which $h_0$ and $n_s$ are free. Parameter entries give the posterior mean
      and marginal $68\%$ standard deviation. The $\chi^2$ rows are evaluated at
      the maximum-posterior sample; ``joint'' uses the full $w(\theta)+\gamma_t$
      covariance, while the $w(\theta)$ and GGL rows use the corresponding
      covariance sub-blocks.}}
      \label{tab:relaxed-marginal-constraints}
      \begin{tabular}{l c cc cc cc}
      \toprule
      Parameter & Truth & \multicolumn{2}{c}{KiDS-like} & \multicolumn{2}{c}{DES-like} & \multicolumn{2}{c}{LSST-like} \\
      \cmidrule(lr){3-4}\cmidrule(lr){5-6}\cmidrule(lr){7-8}
      &  & Uncontam. & Contam. & Uncontam. & Contam. & Uncontam. & Contam. \\
      \midrule
      $\Omega_m$ & $0.3153$ & $0.3296\pm0.0410$ & $0.3342\pm0.0407$ & $0.3234\pm0.0312$ & $0.3305\pm0.0320$ & $0.3167\pm0.0151$ & $0.3275\pm0.0154$ \\
      $\sigma_8$ & $0.8113$ & $0.7900\pm0.0779$ & $0.7823\pm0.0753$ & $0.8012\pm0.0585$ & $0.7869\pm0.0574$ & $0.8100\pm0.0275$ & $0.7902\pm0.0263$ \\
      $S_8$ & $0.8317$ & $0.8219\pm0.0424$ & $0.8199\pm0.0412$ & $0.8282\pm0.0260$ & $0.8222\pm0.0256$ & $0.8314\pm0.0115$ & $0.8247\pm0.0111$ \\
      $h_0$ & $0.6736$ & $0.6634\pm0.0817$ & $0.6657\pm0.0823$ & $0.6678\pm0.0764$ & $0.6619\pm0.0750$ & $0.6770\pm0.0677$ & $0.6544\pm0.0691$ \\
      $n_s$ & $0.9649$ & $0.9622\pm0.0567$ & $0.9641\pm0.0567$ & $0.9633\pm0.0561$ & $0.9600\pm0.0556$ & $0.9635\pm0.0511$ & $0.9873\pm0.0520$ \\
      $\Omega_b$ & $0.0493$ & $0.0505\pm0.0108$ & $0.0503\pm0.0107$ & $0.0496\pm0.0099$ & $0.0498\pm0.0100$ & $0.0491\pm0.0072$ & $0.0502\pm0.0077$ \\
      {$10^9 A_s$} & $2.10$ & $2.050\pm0.433$ & $1.946\pm0.400$ & $2.096\pm0.358$ & $2.008\pm0.340$ & $2.123\pm0.338$ & $2.074\pm0.334$ \\
      $b_1$ & $1.50$ & $1.549\pm0.142$ & $1.568\pm0.138$ & $1.525\pm0.100$ & $1.552\pm0.102$ & $1.504\pm0.046$ & $1.549\pm0.047$ \\
      $b_2$ & $1.80$ & $1.860\pm0.163$ & $1.881\pm0.159$ & $1.830\pm0.116$ & $1.862\pm0.117$ & $1.804\pm0.052$ & $1.856\pm0.052$ \\
      $b_3$ & $1.80$ & $1.861\pm0.166$ & $1.891\pm0.163$ & $1.830\pm0.116$ & $1.872\pm0.119$ & $1.805\pm0.050$ & $1.869\pm0.051$ \\
      $b_4$ & $1.90$ & $1.963\pm0.175$ & $2.026\pm0.174$ & $1.931\pm0.123$ & $2.009\pm0.127$ & $1.905\pm0.053$ & $2.024\pm0.055$ \\
      $b_5$ & $2.30$ & $2.377\pm0.209$ & $2.492\pm0.210$ & $2.337\pm0.148$ & $2.472\pm0.155$ & $2.306\pm0.064$ & $2.505\pm0.068$ \\
      $b_6$ & $2.30$ & $2.377\pm0.209$ & $2.486\pm0.210$ & $2.337\pm0.147$ & $2.465\pm0.154$ & $2.306\pm0.064$ & $2.492\pm0.067$ \\
      \midrule
      {$\chi^2_{\rm joint}$} & {--} & {$0.681$} & {$1.106$} & {$0.873$} & {$1.331$} & {$1.046$} & {$25.099$} \\
      {$\chi^2_{\rm GGL}$} & {--} & {$0.189$} & {$0.164$} & {$0.159$} & {$0.433$} & {$0.420$} & {$6.961$} \\
      {$\chi^2_{w(\theta)}$} & {--} & {$0.525$} & {$1.045$} & {$0.675$} & {$1.041$} & {$0.732$} & {$18.003$} \\
      \bottomrule
      \end{tabular}
      \end{table*}

      \section{Dynamical dark energy}
      \label{sec:app_w0wa}

      The analyses of Sect.~\ref{sec:cosmo} and
      Appendix~\ref{sec:app_relaxed_cosmology} assume a $\Lambda$CDM
      background, in which the redshift-distribution contamination projects
      onto the present-day clustering amplitude ($\sigma_8$, $S_8$) and the
      galaxy-bias parameters. Because the excess clustering is concentrated in
      the high-redshift lens bins (Sect.~\ref{sec:main_results}), it alters the
      redshift dependence of the signal and could therefore be partly
      reinterpreted as a change in the low-redshift growth history rather than
      as a shift in the amplitude. To test this, we extend the inference to a
      dynamical dark-energy ($w_0w_a$CDM) model, in which the dark-energy
      equation of state follows the CPL parametrization
      $w(a)=w_0+w_a(1-a)$ \citep{Chevallier2001, Linder2003}.

      We start from the relaxed configuration of
      Appendix~\ref{sec:app_relaxed_cosmology} ($h_0$ and $n_s$ free) and
      additionally free $w_0$ and $w_a$, with flat priors
      $w_0\in[-3,-1/3]$ and $w_a\in[-3,3]$ and the physical restriction
      $w_0+w_a<0$, {which makes the dark-energy density negligible relative to matter at sufficiently early times};
      the phantom-crossing regime is handled with the parameterized
      post-Friedmann prescription \citep{Fang2008} in \textsc{camb}. The data
      vectors, covariance, likelihood, and scale cuts are identical to the
      $\Lambda$CDM chains, and as before we compare fits to the uncontaminated
      and contaminated data vectors for the three survey configurations.

\begin{figure*}
            \centering
            \includegraphics[width=1\linewidth]{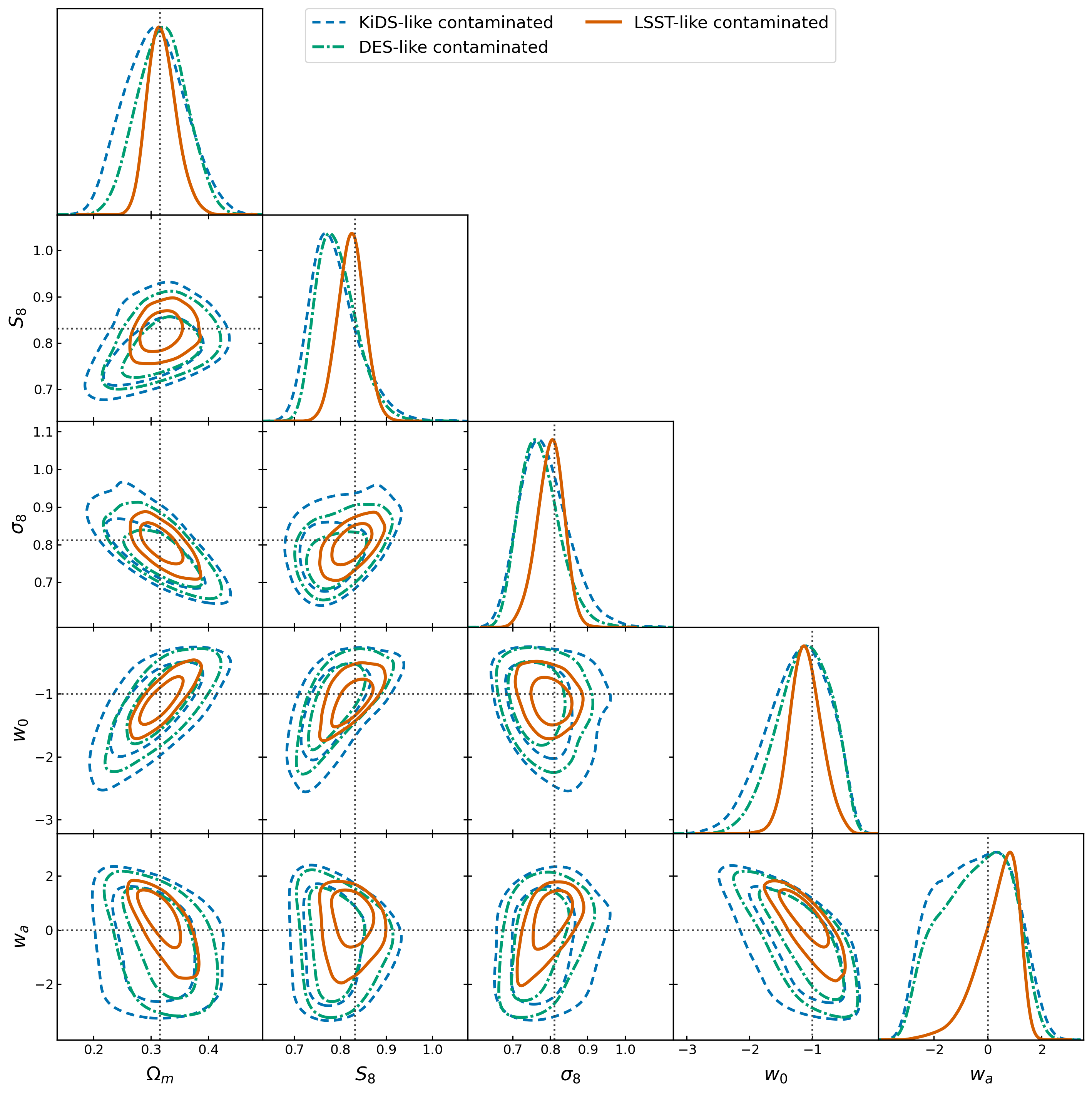}
            \caption{Joint posterior of the matter density $\Omega_m$, the
            amplitude parameters $S_8$ and $\sigma_8$, and the dark-energy
            equation-of-state parameters $w_0$ and $w_a$ in the $w_0w_a$CDM
            extension of Appendix~\ref{sec:app_relaxed_cosmology}, for the
            \emph{contaminated} data vectors of the three survey configurations:
            KiDS-like (blue dashed), DES-like (green dash-dotted), and LSST-like
            (orange solid). Contours show the $68\%$ and $95\%$ credible regions;
            dotted lines mark the fiducial $\Lambda$CDM input values, including
            the cosmological constant $(w_0,w_a)=(-1,0)$. Even with
            contamination all three posteriors recover the fiducial amplitude
            ($S_8$, $\sigma_8$); the only appreciable displacement is in the
            dark-energy sector of the most constraining, LSST-like
            configuration, whose posterior is pulled toward $w_a>0$ and $w_0<-1$
            and is limited there by the $w_0+w_a<0$ prior boundary.}
            \label{fig:w0wa}
      \end{figure*}

      Figure~\ref{fig:w0wa} shows the joint posteriors of the matter density
      $\Omega_m$, the amplitude parameters $S_8$ and $\sigma_8$, and the
      equation-of-state parameters $w_0$ and $w_a$ for the contaminated data
      vectors of the three surveys. Photometric galaxy clustering and galaxy--galaxy
      lensing at these scales constrain the dark-energy equation of state only
      weakly, so the $w_0$/$w_a$ contours are broad and, for the contaminated
      LSST-like case, lean against the $w_0+w_a<0$ prior boundary; the
      dark-energy constraints should therefore be read qualitatively. Even in
      the presence of contamination, all three posteriors recover the fiducial
      amplitude: the $S_8$ and $\sigma_8$ contours remain centered on the input
      values (dotted lines).

      The contamination displaces the $w_0$--$w_a$ posterior, and, as for the
      amplitude parameters in the $\Lambda$CDM chains, the effect grows with
      survey statistical power. For the KiDS-like and DES-like configurations
      the shifts are well within the statistical uncertainty
      ($|\Delta w_0|, |\Delta w_a|\lesssim0.15\sigma$). For the LSST-like
      configuration the equation of state moves toward the late-accelerating quadrant:
      $w_0$ shifts from $-0.94$ to $-1.10$
      ($-0.6\sigma$) and, more markedly, $w_a$ shifts from $-0.46$ to $+0.35$
      ($+1.0\sigma$).

      {The shifts in $\sigma_8$ and $S_8$ are much smaller than in the
      $\Lambda$CDM analysis and are close to zero.} In the relaxed
      $\Lambda$CDM LSST-like chain the contamination shifts $\sigma_8$ by
      $-0.75\sigma$ and $S_8$ by $-0.60\sigma$
      (Appendix~\ref{sec:app_relaxed_cosmology}){. Once} $w_0$ and $w_a$ are
      free, the corresponding shifts fall to $+0.18\sigma$ and $+0.06\sigma$
      ($\sigma_8: 0.794\to0.800$, $S_8: 0.823\to0.824$). The high-redshift
      clustering excess is thus absorbed almost entirely by the dark-energy
      {parameter} rather than by the amplitude{.} {{Freeing} the growth history lets the
      model reinterpret the excess as a mild late-time evolution of $w$,
      leaving $\sigma_8$ and $S_8$ essentially unbiased.} {The latter is} also partly
      due to reduced significance of $\sigma_8$ and $S_8$ measurements with
      relaxed dark energy parameters.

      This confirms the expectation raised in
      Sections~\ref{sec:cosmo_cosmo} and \ref{sec:cosmo_survey}: for a Stage-IV survey, an unmodeled
      anisotropic $n(z,\boldsymbol{\theta})$ can manifest as dynamical dark
      energy, producing a spurious $\sim1\sigma$ preference for $w_a>0$ (and
      $w_0<-1$) rather than a low-$\sigma_8$ signal. {Although the significance here is modest and limited by the priors,
      the result shows that redshift-distribution systematics can affect the
      inferred dark-energy evolution as well as the clustering amplitude.
      This degeneracy must be controlled when using photometric clustering
      to test the dark-energy equation of state.}

      \section{Internal consistency for the DES-like configuration}
      \label{sec:app_ppd_des}

      In the main text, we showed that leaving the modeled systematics unaccounted
      for makes the clustering and lensing observables inconsistent at LSST-like
      statistical precision. Because the significance of this inconsistency depends
      strongly on the data covariance, it is important to test whether the effect
      remains negligible for a Stage-III survey. {Related discrepancies have appeared in DES analyses. After unblinding,
      DES Y3 removed the two highest-redshift MagLim lens bins due to poor fits
      \citep{DESY32022}. DES Y6 removed lens bin 2 before completing unblinding,
      following failed consistency tests \citep{DESY62026}.}
      Following the results of our study, spatial systematics are one candidate for causing such a discrepancy.
  {{Therefore} we test whether spatial variation in $n(z)$ can indeed account
      for the internal inconsistency found by DES.}

\begin{figure}
         \centering
         \includegraphics[width=\linewidth]{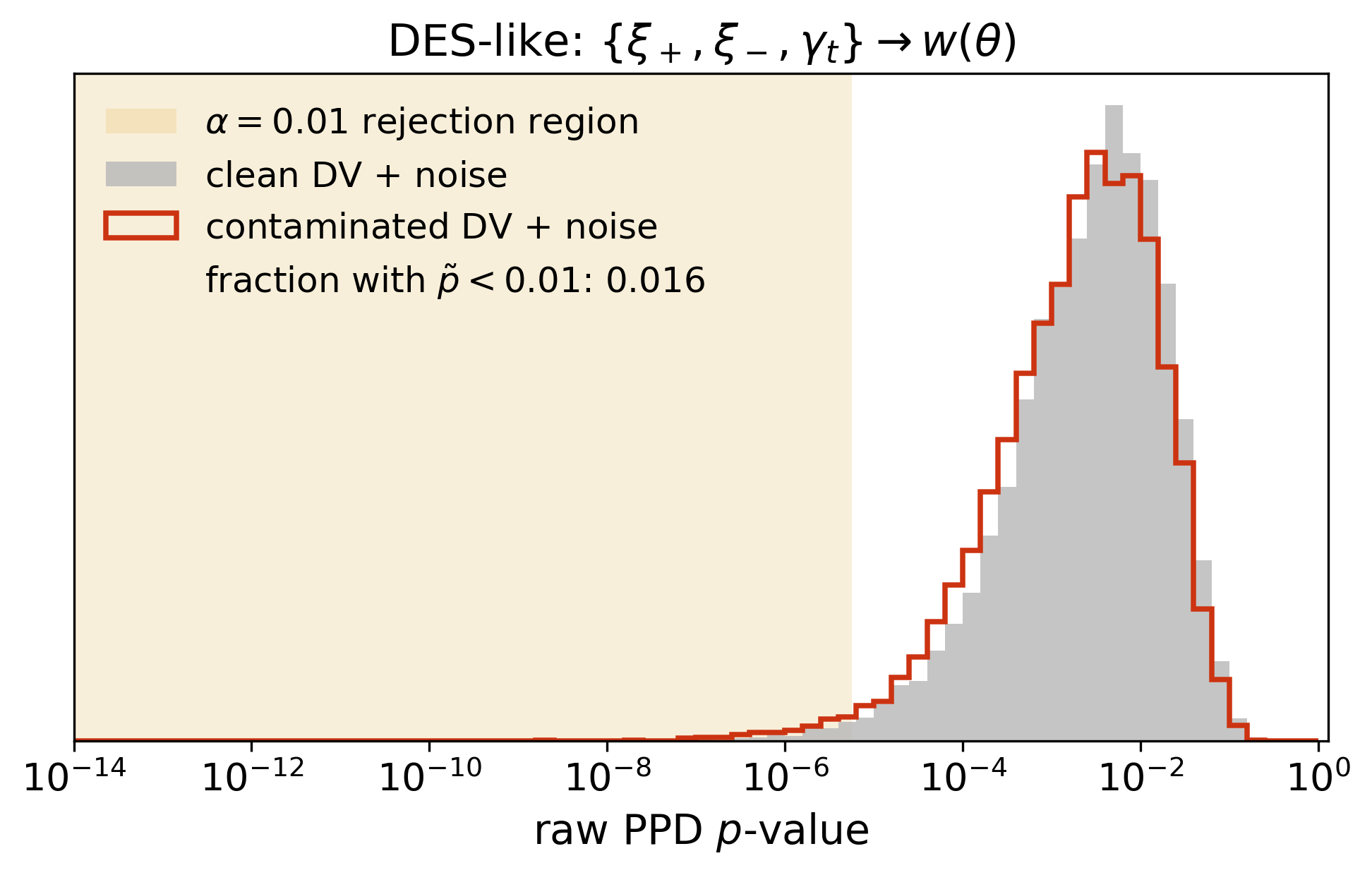}
         \caption{Sampling distributions of the raw PPD $p$-value for the DES-like
         configuration, conditioning on cosmic shear and GGL and predicting
         $w(\theta)$. Gray shows $10^4$ consistent realizations; red shows the same
         realizations with the $w(\theta)$ contamination added. The shaded region is
         the $\alpha=0.01$ rejection region.}
         \label{fig:ppd_consistency_des}
      \end{figure}

      We therefore repeat the PPD analysis for the DES-like configuration. Relative to
      the LSST-like setup, the smaller tile size shifts the contamination to somewhat
      smaller angular scales. Its amplitude remains comparable because the
      observing-condition maps and selection emulator are held fixed to be KiDS-like.
      The DES-like survey area and galaxy number densities enter through the
      covariance and largely determine the sensitivity of this test.
      The results are presented in Fig. \ref{fig:ppd_consistency_des}.
      The median $\tilde{p}$-value is 0.44 {and only 1.6\% of realizations have $\tilde p<0.01$},
  indicating the contamination barely moves the $p$-value given consistent data vectors.
  {The systematics in DES may differ from those modeled here.
  Nevertheless, these results disfavor spatial variation in $n(z)$ as the main
  source of the DES internal inconsistency under our modeling assumptions.}

   \end{appendix}
\end{document}